\documentclass[aps,cp,amsmath,amssymb,reprint,showkeys,showpacs]{revtex4}
\usepackage{graphicx}% Include figure files
\usepackage{dcolumn}% Align table colum
\usepackage{bm}% bold math
\usepackage[utf8]{inputenc}
\usepackage[T1]{fontenc}
\usepackage{mathptmx} 
\usepackage{epsfig}
\usepackage{float}  
\usepackage{xcolor}
\usepackage{soul}  
\newcommand{\eq}{\begin{equation}}
\newcommand{\eeq}{\end{equation}}
\newcommand{\beqa}{\begin{eqnarray}}
\newcommand{\eeqa}{\end{eqnarray}}

\def\ii{\'i}

\begin{document}
\title{Collective description of low energy mesonic states}
\author{Tochtli Y\'epez-Mart\'{\i}nez}
\affiliation{Instituto de Educación Media Superior de la Ciudad de México, 
Plantel General Lázaro Cárdenas del Río, 
Av. Jalalpa Norte 120, C.P. 01377, and,
Instituto Tecnológico de Álvaro Obregón
Calle del Rosal 89, Col. Tepeaca, C.P. 01550, Ciudad de Mexico, Mexico,}

\author{O. A. Rico-Trejo}
\affiliation{Instituto de Ciencias Nucleares, Universidad Nacional
  Aut\'onoma de M\'exico, 
Ciudad Universitaria,
Circuito Exterior S/N, A.P. 70-543, 04510 Ciudad de Mexico, Mexico,}

\author{Peter O. Hess}
\affiliation{Instituto de Ciencias Nucleares, Universidad Nacional
  Aut\'onoma de M\'exico, 
Ciudad Universitaria,
Circuito Exterior S/N, A.P. 70-543, 04510 Ciudad de Mexico, Mexico,
and,
Frankfurt Institute for Advanced Studies, J. W. von Goethe University, Hessen, Germany}
\author{Osvaldo Civitarese}
\affiliation{Departamento de F\'isica, Universidad Nacional de La Plata,
49 y 115. C.C. 67 (1900), La Plata, Argentina,
and, IFLP-CONICET,
diag 115 y 64. La Plata, Pcia Bs.As. Argentina}

\date{\today} % It is always \today, today, but any date may be explicitly specified
              % Not printed for conference proceedings

\begin{abstract}
The phenomenology of hadronic states at high energy is well described in the framework of 
Quantum Chromodynamics. The theory, well established by now, cannot be applied to the description of quark-antiquark 
states at low energy unless their degrees of freedom are treated as effective ones, a condition resulting from confinement. Own to the similarities with other quantum many-body systems we have adopted a non-perturbative scheme where quarks are treated as quasiparticles, which further interact between them, allowing meson-like states to be described as collective superposition of quasiparticle-pairs. The results of the calculations, performed in the context of the random phase approximation method, show that this scheme is a suitable one to describe meson states up to  energies of the order of couple of GeV.
\end{abstract}


\maketitle
\section{Introduction}
\label{intro}

The description of the low energy hadronic spectrum by means of elementary degrees of freedom, that 
is quarks and gluons, is matter of intensive research, particularly, because of the 
non-perturbative regime of Quantum Chromodynamics (QCD) in that energy 
domain \cite{qcd1,qcd2}. QCD is the theory of the strong interactions. It is well understood at
high energies where, due to asymptotic freedom, its coupling-terms are small and 
therefore the various methods of perturbation theory can be applied.

In the low-energy regime QCD is a highly non-perturbative theory, and therefore the description of observables, like the meson spectra, requires the use of another type of techniques. 
Lattice Gauge Theory (LGT) \cite{qcd3}-\cite{latt3} has provided significant results for physical states belonging to 
the low-energy domain \cite{latt4}. 
However, the implementation  of LGT requires huge numerical effort. 
Results, obtained by applying LGT on the regime 
of highly excited meson states, have been reported in Refs.\cite{latt5,latt6},

Among other non-perturbative techniques, the Dyson-Schwinger equation (DSE) method has been applied to low energy QCD \cite{dse1,dse2}. The DSE procedure has its own disadvantages, such as the lack of
clear prescriptions about how to  truncate diagrammatic expansions. 

The formulation of QCD in the Coulomb gauge
\cite{heff1}-\cite{heff8} has been widely used. The calculations for systems involving heavy quarks
have reproduced, approximately, the available data \cite{chino}.
 Recently, we have investigated the
implementation of many-body methods in a $SO(4)$-model for the quark
sector of the QCD Hamiltonian \cite{so44}-\cite{so43}, 
which leads to a pion-like state with high collectivity.

In order to explore the possibilities which are available to formulate the theoretical description of the meson spectra at low energies, one may look at other many body systems where the use of effective degrees of freedom facilitates the calculation of energy spectra and decay rates.
Among the techniques which have been applied to quantum many body systems are the like particle and quasiparticle 
mean field approaches, the Tamm-Dancoff and the Random Phase Approximation 
collective methods \cite{nuc2,nuc1}. 

The starting point of all these methods is the identification of effective degrees of freedom, which in this case are fermionic degrees of freedom, i.e, effective quark degrees of freedom identified with the eigenstates of a central 
potential resulting from the superposition of the QCD interactions.

As we have shown in our previous publications \cite{base1,base2,pre3} in order to build such a description one may start from the diagonalization of quarks states in the central potential, expressing the wave functions in a harmonic oscillator basis \cite{base3}.
This procedure accounts for the gluon exchange among quarks. 
The eigenvalues are thus taken as effective one-quark energies.

In this work we are developing a scheme based on the successive application of these concepts 
starting from effective quarks degrees of freedom up to the interactions among them.

The first step, as we are going to show in this work, consists of the transformation to 
the quasiparticle-quark basis by means of the Bogoliubov`s method, which accounts for the short range 
attractive quark-pair interactions \cite{qcdbcs1}-\cite{qcdbcs7}. 

This could be taken as a zeroth order description of 
meson states, since mesons can be represented by pairs of quasi-quark states with the proper 
quantum numbers determined by the coupling of the quarks quantum numbers. 

The remaining interactions among pairs of quasi-quarks will be treated in the context of the 
Random Phase Approximation (RPA), by taking the pairs of quasi-quark  states as 
components of the basis where the equations resulting from the linearization method are solved.

In addition to the calculation of the spectra of mesons, in the mass region up to 2.0 GeV, we shall calculate 
the spreading width of the states by using the energy states obtained by solving the RPA equations, and by adding to it a residual interaction between the states. This is done by applying the procedure described in \cite{nuc2}, taking the strength of the residual interactions as a parameter. The actual value of this interaction was found to be of the order of  70 MeV, as we shall show in Section \ref{results}. 

The work is organized as follows. The details of the formalisms are presented in Section \ref{sec:theory},
where the steps leading to the composition of each meson subspace are discussed. The results of the calculations are presented in Section \ref{results}. 
The conclusions are given in Section \ref{conclusions}.

\section{Theory}\label{sec:theory}
The method which we have applied to calculate the spectrum of mesons states with masses up 
to 2.0 GeV consists of the following steps:
\begin{itemize}
\item{Building of the effective quark basis: This is done by performing a pre-diagonalization of the confinement potential for  quark and antiquark states in a harmonic oscillator basis, as a function of the number of shells. The procedure has been presented in \cite{base1}}
\item{Transforming the kinetic and Coulomb term of the Hamiltonian to the quasi-quark basis, by means of the Bogoliubov transformations and associated BCS procedure, to define quasi-quark energies, gaps and occupation factors, for each of the subspaces given in terms of the complete set of quark degrees of freedom. At this point the zeroth order mesonic states can be constructed as superposition of uncorrelated pairs of quasi-quarks \cite{pre3,pre1}}
\item{Solving the lineal equations of the remaining terms of the Hamiltonian, consisting of the creation and aniquilation 
of one and two pairs of coupled quasi-quarks, by applying the Random Phase Approximation method \cite{nuc2} }
\end{itemize}
In the next subsections we shall explain the details of each of these steps.

\subsection{The effective Hamiltonian}
\label{Heffec}

At low energy the effects of the exchange of gluons between quark states can be represented by the interaction 
$V(| {\bf x} -{\bf y} |)=-\frac{\alpha_C}{| {\bf x} -{\bf y}|}+\beta_L |{\bf x} -{\bf y}| $,
which is obtained from a self-consistent treatment of the
interaction between color charge-densities. 
Therefore, in that energy domain, the effective QCD Hamiltonian can be written as 
\beqa\label{Eff-QCD-H}
{H}^{QCD}_{eff} &=&
\int\left\{
{\psi}^{\dagger}({ x})
(-i{\alpha}\cdot{\nabla}+\beta{m})
{ \psi}({\bf x})\right\}d{ x}
-\frac{1}{2}
\int\rho_{c}({\bf x})
V(|{\bf x}-{\bf y}|)\rho^{c}({\bf y})d{\bf x}d{\bf y}
\nonumber \\
& =&  {   K} + {   H}_{{\rm Coul}}~,
\label{eq2}
\eeqa
where we have restricted ourselves to the quark sector of the theory, as said before. Consequently, 
$\rho^c({\bf x}) = \psi^\dagger({\bf x}) T^c \psi({\bf x})$
now represents the quark and antiquark charge density.
In the last line of Eq.(\ref{Eff-QCD-H}) the first term is the kinetic energy,
while the second term is the QCD Instantaneous color Coulomb Interaction (QCD-IcCI)  in its simplified form.
The potential term is the short-range interaction of two quarks with
opposite (color) charges is similar to the classical attractive Coulomb interaction
\cite{heff2,qcdbcs5}. 

We start building the Hamiltonian by introducing the kinetic energy term, for effective quark ($  { b}^\dag,b$) and anti-quark ($  { d},d$) degrees of freedom, after the pre-diagonalization procedure described in \cite{pre1}
(See also Appendix A.1). The basis consists of harmonic oscillator states with principal quantum numbers up to $N_{{\rm cut}}$.  

The renormalization of the quark masses and the parameters of the effective linear plus Coulomb interaction, $\alpha_C$ and $\beta_L$, depend on the cut-off $N_{{\rm cut}}$. These values can be further adjusted, numerically, in order to work in a basis with a smaller number of oscillator shells, $N_0$, but 
the low-energy observables should not change when the cut-off is changed. In order to illustrate  this,
let us discuss the renormalization of the masses and  the renormalization of the interaction parameters
$\alpha_C$ and $\beta_L$.

The mass parameters are contained implicitly in  the single-particle energies. 
The renormalization of the quark masses can be achieved by introducing a 
multiplicative factor, such as
\beqa
m^{f}_{N_{{\rm cut}}} & \approx & 
m^{f}_{N_0} \sqrt{x^f_{{\rm N}_{{\rm cut}}}},
\label{masses}
\eeqa
where we have taken into account that the multiplicative factor may depend on the
flavor of the quark, $q$ = u,d or s.
Concerning the renormalization of the interaction, the linear term  
can be rewritten as
\beqa
\beta_L(N_{{\rm cut}})
\langle {   r} \rangle_{N_{{\rm cut}}} & = & \beta_L(N_{{\rm cut}}) \left(
1 + \frac{\langle {   r}\rangle_{N_{{\rm cut}}>N_0}}{\langle {   r}\rangle_{N_0}}
\right)
\langle {   r}\rangle_{N_0}
\nonumber \\
& = & \beta_L(N_0)\langle {   r}\rangle_{N_0},
\label{linear}
\eeqa
hence the relation between $\beta_L(N_0)$ and $\beta_L(N_{\rm cut})$ is
\beqa
\beta_L(N_{{\rm cut}}) & = & \beta_L(N_0) / \sqrt{x_{N_{{\rm cut}}}}
\nonumber \\
\sqrt{x_{N_{{\rm cut}}}} & = &  \left(
1 + \frac{\langle {   r}\rangle_{N_{{\rm cut}}>N_0}}{\langle {   r}\rangle_{N_0}}
\right).
\label{beta}
\eeqa
A similar argument applies to the Coulomb interaction $\frac{1}{{   r}}$ 
and assuming that the relation $<\frac{1}{{   r}}>$ $\approx$ $\frac{1}{<{   r}>}$ 
is fulfilled, we have $\alpha_C(N_{{\rm cut}}) \approx \alpha_C(N_{0})\sqrt{x_{N_{{\rm cut}}}}$.

The effective kinetic energy acquires the 
following structure
\beqa
K=\sum_{k \pi \gamma} \varepsilon_{k\pi \gamma}\sum_{\mu} 
\left( 
  { b}^\dag_{k\pi\gamma\mu}
  { b}^{k\pi\gamma\mu}
-
  { d}^{k\pi\gamma\mu}
  { d}^\dag_{k\pi\gamma\mu}
\right),
\label{kinetic}
\eeqa
being the quantities $\varepsilon_{k\pi \gamma}$ the eigenvalues to be associated to the effective quark energies.

The pair-like terms of the Hamiltonian, written in the same basis, acquire the form
\beqa
\mathcal{F}_{1, 2;~\Gamma_0, \mu_0 }
&=&
\frac{1}{\sqrt{2}} \bigg\{ \delta_{\lambda_1,\frac{1}{2}} \delta_{\lambda_2,\frac{1}{2}}
\left[ 
{   b}^\dagger_{ \pi_1 k_1 J_1 Y_1 T_1 }  \otimes  {   b}_{  \pi_2  k_2 J_2 \bar{Y}_2 T_2 }
\right]^{L, (11),00}_{M_L, C ,0}
-
\delta_{\lambda_1,-\frac{1}{2}} \delta_{\lambda_2,-\frac{1}{2}}
\left[ {   d}_{ \pi_1 k_1 J_1 Y_1 T_1 }   \otimes   {   d}^\dagger_{ \pi_2  k_2 J_2 \bar{Y}_2 T_2 }
\right]^{L (11)00}_{M_L, C, 0} \bigg\}
\nonumber\\
\mathcal{G}_{1, 2;~\Gamma_0,\mu_0}
&=& \frac{1}{\sqrt{2}}  \bigg\{
\delta_{\lambda_1,-\frac{1}{2}} \delta_{\lambda_2,\frac{1}{2}}
\left[   {   d}_{ \pi_1 k_1 J_1 Y_1 T_1 }   \otimes  {   b}_{  \pi_2  k_2 J_2 \bar{Y}_2 T_2 }
\right]^{L, (11),00}_{M_L, C, 0}
-
\delta_{\lambda_1,\frac{1}{2}} \delta_{\lambda_2,-\frac{1}{2}}
\left[ {   b}^\dagger_{ \pi_1 k_1 J_1 Y_1 T_1 }  \otimes  {   d}^\dagger_{  \pi_2  k_2 J_2 \bar{Y}_2 T_2 }
\right]^{L, (11),00}_{M_L,C, 0}     \bigg\},\nonumber\\
\label{pairs}
\eeqa
where we have used the short-hand notations
$1=\lambda_1\pi_1 k_1 J_1 Y_1 T_1 $ (similarly for the index $2$),
for the quantum numbers of the  intermediate coupling in the interaction $\Gamma_0= L,(11),00$, and for
their magnetic projections $\mu_0=M_L,C,0$, respectively.

With these terms, the Coulomb interaction is rewritten as
\beqa
  {H}_{Coul} &=& - \frac{1}{2} \sum_{L }
\sum_{ \lambda_i \pi_i  k_i J_i  Y_i T_i  } 
V^{L}_{ \{\lambda_i  \pi_i  k_i  J_i Y_i T_i \} } 
\Big(   \left[\mathcal{F}_{12;\Gamma_0}
\mathcal{F}_{34;\bar \Gamma_0}\right]^{\tilde 0}_{\tilde 0}
+\left[\mathcal{F}_{12;\Gamma_0}\mathcal{G}_{34;\bar \Gamma_0}\right]^{\tilde 0}_{\tilde 0}
+\left[\mathcal{G}_{12;\Gamma_0}\mathcal{F}_{34;\bar \Gamma_0}\right]^{\tilde 0}_{\tilde 0}
+\left[\mathcal{G}_{12;\Gamma_0}\mathcal{G}_{34;\bar \Gamma_0}\right]^{\tilde 0}_{\tilde 0}
\Big)~,\nonumber\\
\label{coulomb}
\eeqa
where the upper index $\tilde 0_U=\{0,(00),00\}$ indicates the total
couplings of the interaction in spin, color and flavor hypercharge and
isospin, while the lower index $\tilde 0_L=\{0,0,0\}$ denotes the corresponding 
magnetic numbers. 

The matrix elements appearing in Eq.(\ref{coulomb}) are given by the expression
\beqa\label{mat}
V^{L}_{ \{\lambda_i \pi_i k_i  J_i Y_i T_i \} } 
&=&\sum_{\tau_i N_i l_i}~ V_{\{N_i l_i J_i\}}^{L}~
\alpha^{J_1,T_1}_{\tau_1(N_1l_1),\lambda_1,\pi_1,k_1}
\alpha^{J_2,T_2}_{\tau_2(N_2l_2),\lambda_2,\pi_2,k_2}
\alpha^{J_3,T_3}_{\tau_3(N_3l_3),\lambda_3,\pi_3,k_3}
\alpha^{J_4,T_4}_{\tau_4(N_4l_4),\lambda_4,\pi_4,k_4} \nonumber\\
&\times &
\delta_{\tau_1\tau_2} \delta_{\tau_3\tau_4}~
\delta_{\pi_1,(-1)^{\frac{1}{2}-\tau_1 + l_1} }
\delta_{\pi_2,(-1)^{\frac{1}{2}-\tau_2 + l_2} }
\delta_{\pi_3,(-1)^{\frac{1}{2}-\tau_3 + l_3} }
\delta_{\pi_4,(-1)^{\frac{1}{2}-\tau_4 + l_4} } \nonumber\\
&\times &
(-1)^{\frac{1}{3}+\frac{Y_1}{2}+T_1}\frac{\sqrt{2T_1+1}}{\sqrt{3}}\delta_{T_2T_1}\delta_{Y_2 Y_1}
~(-1)^{\frac{1}{3}+\frac{Y_3}{2}+T_3}\frac{\sqrt{2T_3+1}}{\sqrt{3}}\delta_{T_4T_3}\delta_{Y_4 Y_3}~.
\eeqa
The quantities ($V_{\{N_i l_i J_i\}}^{L}$) are analytic and 
easy to compute in the harmonic oscillator basis, being $\alpha$ the amplitudes in that basis. 
Their explicit expressions are listed in the Appendix (A.1) 

\subsection{Transformation to the quasiparticle basis}
The pair terms of Eq.(\ref{pairs}) are written in the basis of quasi-particle operators by means of the Bogoliubov 
transformations \cite{nuc2,nuc1} for the creation 
\beqa
{   B}^\dag_{k\pi, \gamma_q \mu_q} 
&=& u_{k\pi, \gamma_q }{   b}^\dag_{k\pi ,\gamma_q \mu_q}
-v_{k\pi, \gamma_q}{   d}_{k\pi ,\gamma_q \mu_q}\nonumber\\
{   D}^{\dag k\pi, \gamma_q \mu_q}
 &=& u_{k\pi, \gamma_q }{    d}^{\dag  k\pi, \gamma_q \mu_q}
+v_{k\pi, \gamma_q }{   b}^{k\pi, \gamma_q \mu_q},
\label{bcs}
\eeqa
and the annihilation of quasi-particle operators
\beqa
{   B}^{k\pi, \gamma_q \mu_q} 
&=& u^*_{k\pi, \gamma_q }{   b}^{k\pi, \gamma_q \mu_q} 
-v^*_{k\pi, \gamma_q }{   d}^{\dag k\pi, \gamma_q \mu_q}\nonumber\\
{   D}_{k\pi, \gamma_q \mu_q} 
&=& u^*_{k\pi, \gamma_q }{   d}_{k\pi,  \gamma_q \mu_q}
+v^*_{k\pi, \gamma_q }{   b}^\dag_{k\pi, \gamma_q \mu_q},
\eeqa
where we have used the short hand notation $\gamma_q=\{J_q C_q
(Y_q,T_q)\}$ and $\mu_q=\{M_{J_q} M_{C_q} M_{T_q}\}$ to denote the quantum numbers of quark and antiquark states

The inverse transformation yields 
\beqa\label{BCS-trans}
{   b}^\dag_{k\pi, \gamma_q \mu_q} 
&=& u^*_{k\pi, \gamma_q }{   B}^\dag_{k\pi, \gamma_q \mu_q}
+v_{k\pi, \gamma_q }{   D}_{k\pi, \gamma_q \mu_q}\nonumber\\
{   d}^{\dag  k\pi, \gamma_q \mu_q} 
&=& u^*_{k\pi, \gamma_q }{    D}^{\dag  k\pi, \gamma_q \mu_q}
-v_{k\pi, \gamma_q }{   B}^{k\pi, \gamma_q \mu_q}.
\eeqa

The next step consists of transforming the kinetic (Eq.(\ref{kinetic})) and Coulomb ( Eq.(\ref{coulomb})) terms of the
Hamiltonian to the quasi-quark basis in order to write and solve the Bogoliubov equations, for occupation factors and gaps, for each of the channels entering in these expressions.
For the transformed kinetic energy term we have, in short hand notation: 
\beqa
K&=&\sum_{k={k \pi \gamma}} \varepsilon_{k}\sum_{\mu} 
\bigg\{
(u_{k}u_{k}-v_{k}v_{k})
(  { B}^\dagger_{k \mu}  { B}^{ k \mu}
+  { D}^{\dagger \ k  \mu}    { D}_{k \mu} )
+2 u_{k}v_{k}
(  { B}^\dag_{k\mu}   { D}^{\dagger \ k\mu} 
+  { D}_{k\mu}    { B}^{k\mu}  )
\bigg\}\nonumber\\
&+&\sum_{k \pi \gamma} \varepsilon_{k\pi \gamma}
(2 v_{k}^2-1)\Omega_{k \pi \gamma},
\eeqa
and for the one pair terms of the Coulomb interaction we found, in the same notation used for $K$, the form:
\beqa
&&H_{02}+H_{20}=\sum_{k \pi \gamma \mu} 
\varepsilon_{k\pi \gamma}~~ 2 u_{k}v_{k}~~
(  { B}^\dag_{k\mu}   { D}^{\dagger \ k\mu}  +  { D}_{k\mu}    { B}^{k\mu}  )\nonumber\\
&&+\frac{1}{2}\sum_{ m_{J_1}m_{J_2}, c_1 c_2, M_{T_1}M_{T_2} } 
\langle J_1 m_1,J_2-m_2\mid LM_L\rangle 
\langle (10)c_1,(01){\bar c}_2\mid (11) C\rangle_1 
\frac{(-1)^{T_1-M_{T_1}}}{\sqrt{2T_1+1}} \delta_{M_{T_1}  M_{T_2}}
(-1)^{J_2-m_{J_2}} (-1)^{\chi_{c_2}}(-1)^{\chi_{f_2}}\nonumber\\
&&\times \sum_{ m_{J_3}m_{J_4}, c_3 c_4, M_{T_3}M_{T_4} } 
\langle J_3 m_3, J_4-m_4\mid L-M_L\rangle 
\langle (10)c_3,(01){\bar c}_4\mid (11) \bar{C}\rangle_1 
\frac{(-1)^{T_3-M_{T_3}}}{\sqrt{2T_3+1}} \delta_{M_{T_3}  M_{T_4}}
(-1)^{J_4-m_{J_4}} (-1)^{\chi_{c_4}}(-1)^{\chi_{f_4}}\nonumber\\
&&\bigg \{
+\delta_{23}
\bigg \{
\delta_{++++}  (u_{k_2} u_{k_3})    u_{k_1}v_{k_4}
-\delta_{++--}  (u_{k_2} v_{k_3})   u_{k_1}u_{k_4}
+\delta_{--++}  (v_{k_2} u_{k_3})    v_{k_1}v_{k_4}
-\delta_{----}  (v_{k_2} v_{k_3})   v_{k_1}u_{k_4}
\bigg\}
  {  B}^\dagger_{k_1 \mu_1}  { D}^{\dagger \ k_4 \mu_4}\nonumber\\
&&-\delta_{14}
\bigg \{
\delta_{++++}  (v_{k_1} v_{k_4})    v_{k_2}u_{k_3}
-\delta_{++--}  (v_{k_1} u_{k_4})    v_{k_2}v_{k_3}
+\delta_{--++}  (u_{k_1} v_{k_4})    u_{k_2}u_{k_3}
-\delta_{----}  (u_{k_1} u_{k_4})    u_{k_2}v_{k_3}
\bigg\}
  { B}^\dagger_{k_3 \mu_3}  { D}^{\dagger \ k_2 \mu_2}\nonumber\\
&&+\delta_{2 3}\bigg \{ 
\delta_{++++} (u_{k_2}u_{k_3})    v_{k_1}u_{k_4} 
+\delta_{++--} (u_{k_2} v_{k_3})  v_{k_1}v_{k_4}
-\delta_{--++} (v_{k_2}u_{k_3})   u_{k_1}u_{k_4}
-\delta_{----} (v_{k_2} v_{k_3})   u_{k_1}v_{k_4}
\bigg \}  {  D}_{k_1 \mu_1}   { B}^{ k_4 \mu_4} \nonumber\\
&&-\delta_{14}\bigg \{ \delta_{++++} (v_{k_1} v_{k_4})   u_{k_2}v_{k_3}
+\delta_{++--}(v_{k_1}u_{k_4})   u_{k_2}u_{k_3}
-\delta_{--++}(u_{k_1} v_{k_4})   v_{k_2}v_{k_3}
-\delta_{----}(u_{k_1}u_{k_4})     v_{k_2}u_{k_3}
\bigg \}
  { D}_{k_3 \mu_3}   { B}^{ k_2 \mu_2} \nonumber\\
&&+\delta_{23}
\bigg \{
\delta_{+++-}
(u_{k_2} u_{k_3})    u_{k_1}u_{k_4}
-\delta_{++-+}
(u_{k_2} v_{k_3})    u_{k_1}v_{k_4}
+\delta_{--+-}
(v_{k_2} u_{k_3})     v_{k_1}u_{k_4}
-\delta_{---+}
(v_{k_2} v_{k_3})    v_{k_1}v_{k_4}\bigg\}
  {  B}^\dagger_{k_1 \mu_1}  { D}^{\dagger \ k_4 \mu_4}\nonumber\\
&&-\delta_{1 4}
\bigg \{
\delta_{+++-}
(v_{k_1} u_{k_4})    v_{k_2}u_{k_3}
-\delta_{++-+}
(v_{k_1} v_{k_4})    v_{k_2}v_{k_3}
% &&
+\delta_{--+-}
(u_{k_1} u_{k_4})    u_{k_2}u_{k_3}
-\delta_{---+}
(u_{k_1} v_{k_4})    u_{k_2}v_{k_3}\bigg\}
  { B}^\dagger_{k_3 \mu_3}  { D}^{\dagger \ k_2 \mu_2}\nonumber\\
&&-\delta_{23}\bigg\{\delta_{+++-} (u_{k_2}  u_{k_3})v_{k_1}v_{k_4}
+\delta_{++-+} (u_{k_2}  v_{k_3})    v_{k_1} u_{k_4}
-\delta_{--+-} (v_{k_2}  u_{k_3})    u_{k_1}v_{k_4}
-\delta_{---+} (v_{k_2}  v_{k_3})    u_{k_1} u_{k_4} 
\bigg\}  {  D}_{k_1 \mu_1}   { B}^{k_4 \mu_4} \nonumber\\
&&-\delta_{14}\bigg\{\delta_{++-+} (v_{k_1}v_{k_4})u_{k_2} u_{k_3}
+\delta_{+++-}(v_{k_1} u_{k_4})    u_{k_2} v_{k_3}
-\delta_{---+}(u_{k_1}v_{k_4})      v_{k_2} u_{k_3}
-\delta_{--+-} (u_{k_1} u_{k_4})    v_{k_2} v_{k_3}
\bigg\}  { D}_{k_3 \mu_3}  { B}^{ k_2 \mu_2}\nonumber\\
&&-\delta_{2 3}
\bigg \{
\delta_{+-++}
(v_{k_2} u_{k_3})    u_{k_1}v_{k_4}
-\delta_{+---}
(v_{k_2} v_{k_3})    u_{k_1}u_{k_4}
+\delta_{-+++}
(u_{k_2} u_{k_3})     v_{k_1}v_{k_4}
-\delta_{-+--}
(u_{k_2} v_{k_3})    v_{k_1}u_{k_4}\bigg\}
  {  B}^\dagger_{k_1 \mu_1}  { D}^{\dagger \ k_4 \mu_4}\nonumber\\
&&-\delta_{1 4}
\bigg \{
\delta_{-+++}
(u_{k_1} v_{k_4})    v_{k_2}u_{k_3}
-\delta_{-+--}
(u_{k_1} u_{k_4})    v_{k_2}v_{k_3}
+\delta_{+-++}
(v_{k_1} v_{k_4})    u_{k_2}u_{k_3}
-\delta_{+---}
(v_{k_1} u_{k_4})    u_{k_2}v_{k_3}\bigg\}
  { B}^\dagger_{k_3 \mu_3}  { D}^{\dagger \ k_2 \mu_2}\nonumber\\
&&+\delta_{23}
\bigg\{\delta_{-+++} (u_{k_2} u_{k_3})    u_{k_1}u_{k_4}
+\delta_{-+--} ( u_{k_2} v_{k_3})   u_{k_1} v_{k_4}
-\delta_{+-++} (v_{k_2}u_{k_3})    v_{k_1} u_{k_4} 
-\delta_{+---} (v_{k_2} v_{k_3})    v_{k_1} v_{k_4} 
\bigg\}  {  D}_{k_1 \mu_1}   { B}^{k_4 \mu_4}\nonumber\\
&&-\delta_{14}\bigg\{\delta_{-+++} (u_{k_1} v_{k_4})     u_{k_2} v_{k_3}
+\delta_{-+--} (u_{k_1}u_{k_4})     u_{k_2} u_{k_3}
-\delta_{+-++} (v_{k_1} v_{k_4})    v_{k_2}v_{k_3}
-\delta_{+---} (v_{k_1} u_{k_4})     v_{k_2}u_{k_3}
\bigg\}  { D}_{k_3 \mu_3}  { B}^{ k_2 \mu_2}\nonumber\\
&&-\delta_{2 3}
\bigg \{
\delta_{+-+-}
(v_{k_2} u_{k_3})    u_{k_1}u_{k_4}
-\delta_{+--+}
(v_{k_2} v_{k_3})    u_{k_1}v_{k_4}
+\delta_{-++-}
(u_{k_2} u_{k_3})     v_{k_1}u_{k_4}
-\delta_{-+-+}
(u_{k_2} v_{k_3})    v_{k_1}v_{k_4}\bigg\}
  {  B}^\dagger_{k_1 \mu_1}  { D}^{\dagger \ k_4 \mu_4}\nonumber\\
&&-\delta_{1 4}
\bigg \{
\delta_{-++-}
(u_{k_1} u_{k_4})    v_{k_2}u_{k_3}
-\delta_{-+-+}
(u_{k_1} v_{k_4})    v_{k_2}v_{k_3}
+\delta_{+-+-}
(v_{k_1} u_{k_4})    u_{k_2}u_{k_3}
-\delta_{+--+}
(v_{k_1} v_{k_4})    u_{k_2}v_{k_3}\bigg\}
  { B}^\dagger_{k_3 \mu_3}  { D}^{\dagger \ k_2 \mu_2}\nonumber\\
&&-\delta_{23}
\bigg\{
\delta_{-++-} (u_{k_2}  u_{k_3})    u_{k_1}v_{k_4}
+\delta_{-+-+} (u_{k_2}  v_{k_3})    u_{k_1} u_{k_4}
-\delta_{+-+-} (v_{k_2}  u_{k_3})    v_{k_1} v_{k_4}
-\delta_{+--+} (v_{k_2}  v_{k_3})    v_{k_1} u_{k_4}
\bigg\}  {  D}_{k_1 \mu_1}   { B}^{k_4 \mu_4}\nonumber\\
&&-\delta_{14}
\bigg\{\delta_{-+-+} (u_{k_1}v_{k_4})     u_{k_2} u_{k_3}
+\delta_{-++-} (u_{k_1} u_{k_4})    u_{k_2}  v_{k_3}
-\delta_{+--+} (v_{k_1} v_{k_4})     v_{k_2} u_{k_3}
-\delta_{+-+-} (v_{k_1} u_{k_4})    v_{k_2} v_{k_3}
\bigg\}  { D}_{k_3 \mu_3}  { B}^{ k_2 \mu_2} \nonumber\\
&&\bigg \}. \nonumber\\
\eeqa
In the above equation the symbols $\delta_{++++}, \delta_{+--+}$, etc, denote the isospin projection of each of the four quasi-quark states and $\delta_{ab}$ denotes the contractions of the indices of any of the two operators $B$ and $D$.

As said before, the BCS procedure requires that the expectation value of the terms with two creation or two annihilation 
quasi-quark operators should vanish when acting on the correlated vacuum. Together with the condition that the term with one quasi-quark creation and annihilation operators should be diagonal when acting on the correlated 
vacuum. These requirements lead to the determination of the values of the gap and occupation factor for each 
quark flavor. 
The resulting set of equations, using the same short hand notation,  is the following:
\beqa
\Sigma_{k_1} V_{k_1}+\Delta_{k_1} W_{k_1} &=&E_{k_1}\nonumber \\
-\Delta_{k_1} V_{k_1} +\Sigma_{k_1} W_{k_1}&=&0,
\label{bcs1}
\eeqa
where $E_{k_1}$ are the quasi-quark energies, for each state with quantum numbers $k={k \pi \gamma} $, and 
\beqa
\Sigma_{k_1}&=&2\varepsilon_{k_2}\delta_{k_1,k_2}
+\bar{V}^\Sigma_{k_1k_2k_2k_1}(u_{k_2}^2-v_{k_2}^2) \nonumber \\
V_{k_1}&=&\bigg\{u_{k_1}^2-v_{k_1}^2\bigg\}\nonumber \\
\Delta_{k_1}&=&\bar{V}^\Delta_{k_1k_2k_2k_1}   (u_{k_2}v_{k_2}) \nonumber \\
W_{k_1}&=&\bigg\{2u_{k_1}v_{k_1}\bigg\}.
\label{bcs2}
\eeqa
From the solution of these coupled equations are extracted the values of the occupation factors $u_k,v_k$ and gaps
$\Delta_{k}$ for each channel. The quantities entering both Eq.(\ref{bcs1})-and Eq.(\ref{bcs2}) are expressed in terms of the
matrix elements
\beqa
&&\bar{V}^\Sigma_{k_1\pi_1J_1Y_1T_1,
~~k_2\pi_2J_2Y_2T_2 , 
~~k_2\pi_2J_2Y_2T_2,
~~ k_1\pi_1J_1Y_1T_1}=\nonumber\\
&&
- \frac{1}{2} ~ 
\sum_{L}  \sum_{\lambda_i}
~\left(\frac{1}{2}\right)\frac{\sqrt{8(2L+1)}}{9} \frac{   (-1)^{L+J_2-J_1} }{ 2J_1+1 }\nonumber\\
&&\times
\bigg\{
\sum_{\tau_i N_i l_i}~ V_{\{N_i l_i J_i\}}^{L}
\alpha^{J_1,T_1}_{\tau_1(N_1l_1),\lambda_1,\pi_1,k_1}
\alpha^{J_2,T_2}_{\tau_2(N_2l_2),\lambda_2,\pi_2,k_2}
\alpha^{J_3,T_3}_{\tau_3(N_3l_3),\lambda_3,\pi_3,k_3}
\alpha^{J_4,T_4}_{\tau_4(N_4l_4),\lambda_4,\pi_4,k_4} \nonumber\\
&&\times
~\delta_{\tau_1\tau_2} \delta_{\tau_3\tau_4}~
\delta_{ \pi_1 , (-1)^{\frac{1}{2}-\tau_1 + l_1} }
\delta_{  \pi_2 , (-1)^{\frac{1}{2}-\tau_2 + l_2} }
\delta_{ \pi_3 , (-1)^{\frac{1}{2}-\tau_3 + l_3} }
\delta_{  \pi_4 , (-1)^{\frac{1}{2}-\tau_4 + l_4} }      \bigg\}\nonumber\\
&&\times
( \delta_{\pi_1 \pi_4}) (\delta_{k_2 k_3}\delta_{\pi_2 \pi_3}) 
( \delta_{J_2 J_3}\delta_{J_1 J_4} )
( \delta_{T_1 T_2}\delta_{T_2 T_3} \delta_{T_3  T_4})
( \delta_{Y_1 Y_2}\delta_{Y_2 Y_3} \delta_{Y_3  Y_4})\nonumber\\
&&\times 2
(\delta_{\lambda_1+\frac{1}{2}}\delta_{\lambda_2+\frac{1}{2}}
\delta_{\lambda_3+\frac{1}{2}}\delta_{\lambda_4+\frac{1}{2}}
-\delta_{\lambda_1-\frac{1}{2}}\delta_{\lambda_2+\frac{1}{2}}
\delta_{\lambda_3+\frac{1}{2}}\delta_{\lambda_4-\frac{1}{2}})_{\mbox{average}},
\eeqa
and

\beqa
&&\bar{V}^\Delta_{k_1\pi_1J_1Y_1T_1,
~~k_2\pi_2J_2Y_2T_2 , 
~~k_2\pi_2J_2Y_2T_2,
~~ k_1\pi_1J_1Y_1T_1}=\nonumber\\
&&
% \bar{V}_{k_1k_2k_2k_1}=
- \frac{1}{2} ~ 
\sum_{L}  \sum_{\lambda_i}
~\left(\frac{1}{2}\right)\frac{\sqrt{8(2L+1)}}{9} \frac{   (-1)^{L+J_2-J_1} }{ 2J_1+1 }\nonumber\\
&&\times
\bigg\{
\sum_{\tau_i N_i l_i}~ V_{\{N_i l_i J_i\}}^{L}
\alpha^{J_1,T_1}_{\tau_1(N_1l_1),\lambda_1,\pi_1,k_1}
\alpha^{J_2,T_2}_{\tau_2(N_2l_2),\lambda_2,\pi_2,k_2}
\alpha^{J_3,T_3}_{\tau_3(N_3l_3),\lambda_3,\pi_3,k_3}
\alpha^{J_4,T_4}_{\tau_4(N_4l_4),\lambda_4,\pi_4,k_4} \nonumber\\
&&\times
~\delta_{\tau_1\tau_2} \delta_{\tau_3\tau_4}~
\delta_{ \pi_1 , (-1)^{\frac{1}{2}-\tau_1 + l_1} }
\delta_{  \pi_2 , (-1)^{\frac{1}{2}-\tau_2 + l_2} }
\delta_{ \pi_3 , (-1)^{\frac{1}{2}-\tau_3 + l_3} }
\delta_{  \pi_4 , (-1)^{\frac{1}{2}-\tau_4 + l_4} }      \bigg\}\nonumber\\
&&\times
( \delta_{\pi_1 \pi_4}) (\delta_{k_2 k_3}\delta_{\pi_2 \pi_3}) 
( \delta_{J_2 J_3}\delta_{J_1 J_4} )
( \delta_{T_1 T_2}\delta_{T_2 T_3} \delta_{T_3  T_4})
( \delta_{Y_1 Y_2}\delta_{Y_2 Y_3} \delta_{Y_3  Y_4})\nonumber\\
&&\times 4
(\delta_{\lambda_1+\frac{1}{2}}\delta_{\lambda_2+\frac{1}{2}}
\delta_{\lambda_3-\frac{1}{2}}\delta_{\lambda_4-\frac{1}{2}}
+\delta_{\lambda_1-\frac{1}{2}}\delta_{\lambda_2+\frac{1}{2}}
\delta_{\lambda_3-\frac{1}{2}}\delta_{\lambda_4+\frac{1}{2}}).
\eeqa
Up to now we have taken into account the pairing-type interaction between quarks by means of the BCS procedure and ended up in the quasi-quark basis. The transformed Hamiltonian has still terms which are acting on the space of  correlated pairs of quasi-quarks and the meson spectrum may be constructed by a new diagonalization by means of the Random Phase Approximation \cite{nuc2}, as it is explained in the next subsection.
\subsection{The Random Phase Approximation procedure}
The Hamiltonian consisting of the kinetic energy (Eq.(\ref{kinetic})) and Coulomb (Eq.(\ref{coulomb})) terms, once transformed to the quasi-quark basis, includes single quasi-quark terms and the creation and annihilation of one and two pairs of coupled quasi-quarks. Let us call this transformed Hamiltonian ${H}^{QCD}_{RPA}$
and its terms $H_{11},H_{22}$ and $H_{40+04}$, respectively. The one pair terms $H_{20+02}$ have been already accounted for by the BCS procedure described in the previous subsection. The Random Phase Approximation method consists of the diagonalization of the Hamiltonian:
\begin{equation}
{   H}^{QCD}_{RPA}=H_{11}+H_{22}+H_{40+04}, 
\label{124}
\end{equation}
in a new basis, the so-called one-phonon basis, where the phonons are lineal combination of quasi-quark pair-creation and quasi-quark pair annihilation operators, which carry the quantum numbers of mesonic states. 
These phonons, $\hat \Gamma^\dagger_{n;\Gamma\mu}$, are defined as:
\begin{equation}
\hat \Gamma^\dag_{n;\Gamma\mu}
=\sum_{{\bf a}, {\bf b}} X^n_{{\bf a}{\bf b};\Gamma}  
{[{   B}^\dag_{\bf a} {   D}^\dag_{\bar {\bf b}} ]}^\Gamma_\mu
-Y^n_{{\bf a}{\bf b};\Gamma} 
(-1)^{\phi_{\Gamma\mu}}
{[{   D}^{\bar {\bf b}}{   B}^{ {\bf a}}  ]}^{\bar \Gamma}_{\bar \mu},
\label{phonon}
\end{equation}
where $(-1)^{\phi_{\Gamma\mu}}=(-1)^{J-M_J}(-1)^{T-T_z}$ is the phase needed to guarantee the scalar nature of the vacuum. 
The  commutator between the one-phonon operator and the Hamiltonian of Eq.(\ref{124}) reads
\begin{equation}
\langle RPA|\left[\hat \Gamma^{n';\Gamma\mu},\left[ {   H}^{QCD}_{RPA} ,\Gamma^\dag_{n;\Gamma\mu}\right]\right]
|RPA \rangle=E^{RPA}_{n;\Gamma} \delta_{n,n'}.
\label{RPA}  
\end{equation}
The quantities $E^{RPA}_{n;\Gamma}$ are the n-th one-phonon eigenvalues to be associated to the mesonic states. 
The $n$-th RPA state with quantum numbers $\Gamma\mu$ is defined as
$| n; \Gamma\mu \rangle_{RPA} =\hat \Gamma^\dag_{n;\Gamma\mu}|RPA\rangle $. 
The RPA ground state  $|RPA\rangle$ is constructed in such a way that the condition
$\hat \Gamma^{n;\Gamma\mu}|RPA\rangle=0$ is fulfilled. 
In terms of the so-called forward- and backward-going amplitudes of the phonon states, $X^n_{{\bf a}{\bf b},\Gamma} $ and 
$Y^n_{{\bf a}{\bf b},\Gamma}$, respectively, the RPA equations of motion , for the pair creation and annihilation terms of the  Hamiltonian (\ref{124}) are written as follows:
\beqa
&&\langle RPA |  
\left[\hat \Gamma^{n';\Gamma\mu},
\left[ H^{QCD}_{22+40+04},
\Gamma^\dag_{n;\Gamma\mu}\right]
\right]|RPA \rangle 
=-\frac{1}{2}\sum_{L}\sum_{\lambda_i \pi_i k_i J_i Y_i T_i }
V^L_{\{ \lambda_i \pi_i k_i J_i Y_i T_i  \}}
\sum_{\mu_0} \frac{(-1)^{\Gamma_0-\mu_0}}{\hat{\Gamma_0}}\nonumber\\
&& \times 
\langle RPA |  
\left[\hat \Gamma^{n';\Gamma\mu},
\left[
\mathcal{F}_{12;\Gamma_0 ~\mu_0}
\mathcal{F}_{34;\bar \Gamma_0 ~\bar \mu_0}+
\mathcal{F}_{12;\Gamma_0 ~\mu_0}
\mathcal{G}_{34;\bar \Gamma_0 ~\bar \mu_0}+
\mathcal{G}_{12;\Gamma_0 ~\mu_0}
\mathcal{F}_{34;\bar \Gamma_0 ~\bar \mu_0}+
\mathcal{G}_{12;\Gamma_0 ~\mu_0}
\mathcal{G}_{34;\bar \Gamma_0 ~\bar \mu_0},
\Gamma^\dag_{n;\Gamma\mu}\right]
\right]|RPA \rangle . \nonumber \\
\label{double} 
\eeqa 
The operators $\mathcal{F}$ and $ \mathcal{G}$  (see Appendix A.2) and their products contain terms with the creation and annihilation of one and two quasiparticle-pairs. The double commutator of the Hamiltonian with the one-phonon operator of Eq.(\ref{phonon}) yields:
\beqa
&&\langle RPA|\left[\hat \Gamma^{n';\Gamma\mu},\left[ {   H}^{QCD}_{22+40+04} ,\Gamma^\dag_{n;\Gamma\mu}\right]\right]
|RPA \rangle=
-\frac{1}{4}\sum_{L}\sum_{\lambda_i \pi_i k_i J_i Y_i T_i }
V^L_{\{ \lambda_i \pi_i k_i J_i Y_i T_i  \}}
\sum_{{\bf a}^{\prime} {\bf b}^{\prime}}
\sum_{{\bf a}{\bf b}}
\left(X^m_{{\bf a}^{\prime}{\bf b}^{\prime};\Gamma^{\prime}}\right)    
\left(X^n_{{\bf a}{\bf b};\Gamma}\right) \nonumber\\
&& \bigg \{ 
 - V^1_{\mathcal{F} \mathcal{F}}~
\delta_{2a}\delta_{3b}
\delta_{1a^{\prime}}\delta_{4b^{\prime}}
 -   V^2_{\mathcal{F} \mathcal{F}}~
\delta_{4a}\delta_{1b}\delta_{3a^{\prime}}\delta_{2b^{\prime}}
 + V^1_{\mathcal{G} \mathcal{G}}~
\delta_{2a}\delta_{3b}
\delta_{1a^{\prime}}\delta_{4b^{\prime}}
 +  V^2_{\mathcal{G} \mathcal{G}}~
\delta_{4a}\delta_{1b}\delta_{3a^{\prime}}\delta_{2b^{\prime}}\nonumber\\
&& -  V^1_{\mathcal{F} \mathcal{G}}~
\delta_{2a}\delta_{3b}
\delta_{1a^{\prime}}\delta_{4b^{\prime}}
 +   V^2_{\mathcal{F} \mathcal{G}}~
\delta_{4a}\delta_{1b}\delta_{3a^{\prime}}\delta_{2b^{\prime}}
 + V^1_{\mathcal{G} \mathcal{F}}\delta_{2a}\delta_{3b}
\delta_{1a^{\prime}}\delta_{4b^{\prime}}
 -  V^2_{\mathcal{G} \mathcal{F}}
\delta_{4a}\delta_{1b}\delta_{3a^{\prime}}\delta_{2b^{\prime}}\bigg \}
\nonumber\\
&& \times  \hat{\Gamma}_0	
(-1)^{\gamma_a+\gamma_b+\Gamma} 
\left\{\begin{array}{lll} 
\bar{\gamma}_{b}^{\prime} & \gamma_{a}^{\prime} & \Gamma \\ [0.01in] 
\bar{\gamma}_{a} &\gamma_{b}&  \Gamma_{0} 
 \end{array}\right\}  
 \delta_{\Gamma \Gamma^{\prime}} \delta_{\mu \mu^{\prime}}
 \nonumber\\
&& - \frac{1}{4}\sum_{L}\sum_{\lambda_i \pi_i k_i J_i Y_i T_i }
V^L_{\{ \lambda_i \pi_i k_i J_i Y_i T_i  \}}
\sum_{{\bf a}^{\prime} {\bf b}^{\prime}}
\sum_{{\bf a}{\bf b}}
\left(X^m_{{\bf a}^{\prime}{\bf b}^{\prime};\Gamma^{\prime}}\right)    \left(-Y^n_{{\bf a}{\bf b};\Gamma}\right)
\nonumber\\
&&\bigg \{ 
 + W^1_{\mathcal{F} \mathcal{F}}~
\delta_{2b}\delta_{3a}
\delta_{1a^{\prime}}\delta_{4b^{\prime}}
 + W^1_{\mathcal{F} \mathcal{F}}~
\delta_{4b}\delta_{1a}\delta_{3a^{\prime}}\delta_{2b^{\prime}}
 + W^1_{\mathcal{G} \mathcal{G}}~
\delta_{2b}\delta_{3a}
\delta_{1a^{\prime}}\delta_{4b^{\prime}}
 + W^1_{\mathcal{G} \mathcal{G}}~
\delta_{4b}\delta_{1a}\delta_{3a^{\prime}}\delta_{2b^{\prime}}\nonumber\\
&&  +  W^1_{\mathcal{F} \mathcal{G}}~
\delta_{2b}\delta_{3a}
\delta_{1a^{\prime}}\delta_{4b^{\prime}}
 + W^1_{\mathcal{F} \mathcal{G}}~
\delta_{4b}\delta_{1a}\delta_{3a^{\prime}}\delta_{2b^{\prime}}
 + W^1_{\mathcal{G} \mathcal{F}}~
\delta_{2b}\delta_{3a}
\delta_{1a^{\prime}}\delta_{4b^{\prime}}
 + W^1_{\mathcal{G} \mathcal{F}}~
\delta_{4b}\delta_{1a}\delta_{3a^{\prime}}\delta_{2b^{\prime}}
~\bigg \}\nonumber\\
&& \times  \hat{\Gamma}_0	
(-1)^{2 \gamma_a} 
\left\{\begin{array}{lll} 
\bar{\gamma}_{b}^{\prime} & \gamma_{a}^{\prime} & \Gamma \\ [0.01in] 
\bar{\gamma}_{b} &\gamma_{a}&  \Gamma_{0} 
 \end{array}\right\}  
 \delta_{\Gamma \Gamma^{\prime}} \delta_{\mu \mu^{\prime}}\nonumber\\
&&-\frac{1}{4}\sum_{L}\sum_{\lambda_i \pi_i k_i J_i Y_i T_i }
V^L_{\{ \lambda_i \pi_i k_i J_i Y_i T_i  \}}
\sum_{{\bf a}^{\prime} {\bf b}^{\prime}}
\sum_{{\bf a} {\bf b}}   
\left(-Y^m_{{\bf a}^{\prime}{\bf b}^{\prime};\Gamma^{\prime}}\right) 
\left(-Y^n_{{\bf a}{\bf b};\Gamma}\right)\nonumber\\
&&\bigg \{
+ V^1_{\mathcal{F} \mathcal{F}}~
\delta_{1a}\delta_{4b}   \delta_{2a^{\prime}}\delta_{3b^{\prime}}
 +   V^2_{\mathcal{F} \mathcal{F}}~
\delta_{3a}\delta_{2b}
\delta_{4a^{\prime}}\delta_{1b^{\prime}}
+ V^1_{\mathcal{G} \mathcal{G}}~
\delta_{1a}\delta_{4b}
\delta_{2a^{\prime}}\delta_{3 b^{\prime}}
 +  V^2_{\mathcal{G} \mathcal{G}}~
\delta_{3a}\delta_{2b}\delta_{4a^{\prime}}\delta_{1b^{\prime}}\nonumber\\
&& + V^1_{\mathcal{F} \mathcal{G}}~
\delta_{1a}\delta_{4b}   \delta_{2a^{\prime}}\delta_{3b^{\prime}}
-  V^2_{\mathcal{F} \mathcal{G}}~
\delta_{3a}\delta_{2b}\delta_{4a^{\prime}}\delta_{1b^{\prime}}
+ V^1_{\mathcal{G} \mathcal{F}}~
\delta_{1a}\delta_{4b}   \delta_{2a^{\prime}}\delta_{3b^{\prime}}
-  V^2_{\mathcal{G} \mathcal{F}}~
\delta_{3a}\delta_{2b}\delta_{4a^{\prime}}\delta_{1b^{\prime}}
~\bigg \}\nonumber\\
&&\times  
  \hat{\Gamma}_0	
(-1)^{\gamma_a^{\prime}+\gamma_b^{\prime}+\Gamma} 
\left\{\begin{array}{lll} 
\bar{\gamma}_{b}^{\prime} & \gamma_{a}^{\prime} & \Gamma \\ [0.01in] 
\bar{\gamma}_{a} &\gamma_{b}&  \Gamma_{0} 
 \end{array}\right\}  
 \delta_{\Gamma \Gamma^{\prime}} \delta_{\mu \mu^{\prime}}
 \nonumber\\
&&  -  \frac{1}{4}\sum_{L}\sum_{\lambda_i \pi_i k_i J_i Y_i T_i }
V^L_{\{ \lambda_i \pi_i k_i J_i Y_i T_i  \}}
\sum_{{\bf a}^{\prime} {\bf b}^{\prime}}
\sum_{{\bf a}{\bf b}}
\left(-Y^m_{{\bf a}^{\prime}{\bf b}^{\prime};\Gamma^{\prime}}\right)    
\left(X^n_{{\bf a}{\bf b};\Gamma}\right)\nonumber\\
&& \bigg \{  + W^2_{\mathcal{F} \mathcal{F}}~
\delta_{2a}\delta_{3b}
\delta_{1b^{\prime}}\delta_{4a^{\prime}}
 + W^2_{\mathcal{F} \mathcal{F}}~
\delta_{4a}\delta_{1b}\delta_{3b^{\prime}}\delta_{2a^{\prime}}
 + W^2_{\mathcal{G} \mathcal{G}}~
\delta_{2a}\delta_{3b}
\delta_{1b^{\prime}}\delta_{4a^{\prime}}
 + W^2_{\mathcal{G} \mathcal{G}}~
\delta_{4a}\delta_{1b}\delta_{3b^{\prime}}\delta_{2a^{\prime}}\nonumber\\
&& +   W^2_{\mathcal{F} \mathcal{G}}~
\delta_{2a}\delta_{3b}
\delta_{1b^{\prime}}\delta_{4a^{\prime}}
 + W^2_{\mathcal{F} \mathcal{G}}~
\delta_{4a}\delta_{1b}\delta_{3b^{\prime}}\delta_{2a^{\prime}}
+ W^2_{\mathcal{G} \mathcal{F}}~
\delta_{2a}\delta_{3b}
\delta_{1b^{\prime}}\delta_{4a^{\prime}}
 + W^2_{\mathcal{G} \mathcal{F}}~
\delta_{4a}\delta_{1b}\delta_{3b^{\prime}}\delta_{2a^{\prime}}
~\bigg \}\nonumber\\
&& \times   \hat{\Gamma}_0	
(-1)^{2 \gamma_b}
(-1)^{\gamma_a+ \gamma_b} (-1)^{\gamma_a^{\prime}+ \gamma_b^{\prime}}
\left\{\begin{array}{lll} 
\bar{\gamma}_{b}^{\prime} & \gamma_{a}^{\prime} & \bar{\Gamma} \\ [0.01in] 
\bar{\gamma}_{b} &\gamma_{a}&  \bar{\Gamma}_{0} 
 \end{array}\right\}  
 \delta_{\Gamma \Gamma^{\prime}} \delta_{\mu \mu^{\prime}}. \nonumber\\
\label{eqrpa}
\eeqa
In the previous expression the symbol
$
\left\{\begin{array}{lll} 
\bar{\gamma}_{b}^{\prime} & \gamma_{a}^{\prime} & \bar{\Gamma} \\ [0.01in] 
\bar{\gamma}_{b} &\gamma_{a}&  \bar{\Gamma}_{0} 
 \end{array}\right\}  $
represents the product of three recoupling coefficients which relates : (i) the coupling of the quantum numbers of a single quasi-quark state, $(k \pi)$, to any of the four values of $\gamma$, (ii) the coupling of pair of quasi-quarks appearing in the operators  $\mathcal{F}$ and $\mathcal{G}$ to $\Gamma_0$, and (iii) the coupling of the quasiparticle pairs to the phonon total angular momentum $\Gamma$.
In the expression of Eq.(\ref{eqrpa}) the factors $V$ and $W$ stand for:
\beqa
&& V^1_{\mathcal{F} \mathcal{F}} =
 (\delta_{++}u_{k_1}u_{k_2} +\delta_{--} v_{k_1} v_{k_2})  ~
(\delta_{++}v_{k_3} v_{k_4}+\delta_{--}u_{k_3}u_{k_4})\nonumber\\
&& V^2_{\mathcal{F} \mathcal{F}} =
(\delta_{++}v_{k_1} v_{k_2}+\delta_{--}u_{k_1}u_{k_2})
(\delta_{++}u_{k_3}u_{k_4} +\delta_{--} v_{k_3} v_{k_4})\nonumber\\
&& W^1_{\mathcal{F} \mathcal{F}}=
(\delta_{++}u_{k_1}v_{k_2}-\delta_{--}v_{k_1}u_{k_2})    
(\delta_{++}u_{k_3}v_{k_4}-\delta_{--}v_{k_3}u_{k_4}) \nonumber\\
&& W^2_{\mathcal{F} \mathcal{F}} =
(\delta_{++}v_{k_1}u_{k_2}-\delta_{--}u_{k_1}v_{k_2})  
(\delta_{++}v_{k_3}u_{k_4}-\delta_{--}u_{k_3}v_{k_4}) 
\eeqa
and
\beqa
&& V^1_{\mathcal{F} \mathcal{G}} =
(\delta_{++}u_{k_1}u_{k_2} +\delta_{--} v_{k_1} v_{k_2})  
(\delta_{-+}u_{k_3} v_{k_4}+\delta_{+-}v_{k_3}u_{k_4})\nonumber\\
&& V^2_{\mathcal{F} \mathcal{G}} =
(\delta_{++}v_{k_1} v_{k_2}+\delta_{--}u_{k_1}u_{k_2})
(\delta_{+-}u_{k_3}v_{k_4} +\delta_{-+} v_{k_3} u_{k_4})\nonumber\\
&& W^1_{\mathcal{F} \mathcal{G}}=
(\delta_{++}u_{k_1}v_{k_2}-\delta_{--}v_{k_1}u_{k_2})    
(\delta_{+-}u_{k_3}u_{k_4}-\delta_{-+}v_{k_3}v_{k_4}) 
\nonumber\\
&& W^2_{\mathcal{F} \mathcal{G}} =
(\delta_{++}v_{k_1}u_{k_2}-\delta_{--}u_{k_1}v_{k_2})  
(\delta_{-+}u_{k_3}u_{k_4}-\delta_{+-}v_{k_3}v_{k_4}) 
\eeqa
\beqa
&& V^1_{\mathcal{G} \mathcal{F}} =
(\delta_{+-}u_{k_1}v_{k_2} +\delta_{-+} v_{k_1} u_{k_2})  
(\delta_{++}v_{k_3} v_{k_4}+\delta_{--}u_{k_3}u_{k_4})\nonumber\\
&& V^2_{\mathcal{G} \mathcal{F}} =
(\delta_{-+}u_{k_1} v_{k_2}+\delta_{+-}v_{k_1}u_{k_2})
(\delta_{++}u_{k_3}u_{k_4} +\delta_{--} v_{k_3} v_{k_4})
\nonumber\\
&& W^1_{\mathcal{G} \mathcal{F}}=
(\delta_{+-}u_{k_1}u_{k_2}-\delta_{-+}v_{k_1}v_{k_2})    
(\delta_{++}u_{k_3}v_{k_4}-\delta_{--}v_{k_3}u_{k_4}) \nonumber\\
&& W^2_{\mathcal{G} \mathcal{F}} =
(\delta_{-+}u_{k_1}u_{k_2}-\delta_{+-}v_{k_1}v_{k_2})  
(\delta_{++}v_{k_3}u_{k_4}-\delta_{--}u_{k_3}v_{k_4}) 
\eeqa
\beqa
&& V^1_{\mathcal{G} \mathcal{G}} =
(\delta_{+-}u_{k_1}v_{k_2} +\delta_{-+} v_{k_1} u_{k_2})  
(\delta_{-+}u_{k_3} v_{k_4}+\delta_{+-}v_{k_3}u_{k_4})\nonumber\\
&& V^2_{\mathcal{G} \mathcal{G}} =
(\delta_{-+}u_{k_1} v_{k_2}+\delta_{+-}v_{k_1}u_{k_2})
(\delta_{+-}u_{k_3}v_{k_4} +\delta_{-+} v_{k_3} u_{k_4})\nonumber\\
&& W^1_{\mathcal{G} \mathcal{G}}=
(\delta_{+-}u_{k_1}u_{k_2}-\delta_{-+}v_{k_1}v_{k_2})    
(\delta_{+-}u_{k_3}u_{k_4}-\delta_{-+}v_{k_3}v_{k_4}) \nonumber\\
&& W^2_{\mathcal{G} \mathcal{G}} =
(\delta_{-+}u_{k_1}u_{k_2}-\delta_{+-}v_{k_1}v_{k_2})  
(\delta_{-+}u_{k_3}u_{k_4}-\delta_{+-}v_{k_3}v_{k_4}) .
\eeqa
The formal solution of the RPA method requires the diagonalization of the coupled equations (\ref{eqrpa}).
Therefore,from it, we get two sets of equations 
which in matrix form can be written:
\beqa\label{RPAcompact}
\left(\begin{array}{c c} A & B  \\B^* & A^*   \\\end{array}\right)
\left(\begin{array}{c} X^n   \\ Y^n    \\ \end{array}\right) 
=E^{RPA}_{n}
\left(\begin{array}{c c} 1 & 0  \\0 & -1   \\\end{array}\right)
\left(\begin{array}{c} X^n   \\ Y^n    \\ \end{array}\right) ~,
\eeqa
with
\beqa\label{Ap:F-and-B}
A_{a'b';\Gamma'\mu';~a,b;\Gamma\mu}&=&
\langle \tilde 0 | \left[ [{   D}^{\bar {\bf b}'}  {   B}^ {{\bf a}'}]^{\Gamma'}_{\mu'},
\left[ {H}^{QCD}_{RPA} ,  [{   B}^\dag_{\bf a} {   D}^\dag_{\bar {\bf b}} ]^\Gamma_\mu \right]  \right] 
| \tilde 0 \rangle ~, \nonumber\\
B_{a'b';\Gamma'\mu';~a,b;\Gamma\mu}&=&-
\langle \tilde 0 | \left[ [{   D}^{\bar {\bf b}'}  {   B}^ {{\bf a}'}]^{\Gamma'}_{\mu'},
\left[ {H}^{QCD}_{RPA} ,(-1)^{\phi_{\Gamma\mu}}
[{   D}^{\bar {\bf b}}  {   B}^{ {\bf a}}  ]^{\Gamma}_{\bar \mu} \right] \right] 
| \tilde 0 \rangle ,
\eeqa
being $A$ and $B$ the forward and backward matrices of the RPA method,
respectively, which are expressed by the terms of Eq.(\ref{eqrpa}) to which are added, in the case of the matrix $A$, the diagonal terms coming from the double commutator with $H_{11}$, which is the Hamiltonian representing the single pair of quasi-quarks. These terms are just the quasiparticle energies of the pair of quasi-quarks. The RPA forward- and backward-going amplitudes are then determined from the solutions of the system of Eq.(\ref{RPAcompact}) with the orthogonalization condition:
\begin{equation}\label{XandY}
\sum_{{\bf a},{\bf b},\Gamma}X^m_{{\bf a},{\bf b},\Gamma}X^n_{{\bf a},{\bf b},\Gamma}   
-Y^m_{{\bf a}{\bf b};\Gamma}Y^n_{{\bf a}{\bf b};\Gamma}=\delta(m,n)
\end{equation}
We have solved the RPA equations for each set of quantum numbers associated to meson states and kept the solutions for eigenvalues up to 2.0 GeV. 
\subsection{Spreading width of the states}
\label{width}
Once the RPA equations are solved we have a set of states of known energies resulting from the treatment of the terms appearing in Eq.(\ref{124}). Let us call, for simplicity $\bf {H}_0$ and $E_a$ the Hamiltonian and the RPA eigenvalues, respectively. Any residual interaction among the states can be 
represented by $\bf {V}$, such that the complete Hamiltonian is written:
\beqa\label{H}
\bf{H}=\bf {H}_0+\bf{V}.
\eeqa
Then, for the effects upon a certain state $| a \rangle$ due to the set of other 
states $| \alpha \rangle$ we write, following Bohr and Mottelson \cite{nuc2}
\beqa\label{elements}
{\bf{H}_0} | a \rangle &=& E_{a} | a \rangle\nonumber\\
{\bf{H}_0} | \alpha \rangle &=& E_{\alpha} | \alpha \rangle\nonumber\\
\langle a | {\bf{V}} | a \rangle &=& 0\nonumber\\
\langle \alpha_j | {\bf{V}} | \alpha_{j'} \rangle &=& V_{ \alpha_{j}, \alpha_{{j'} }}=0
~~~~\forall ~j,j' \nonumber\\
\langle a | {\bf{V}} | \alpha_{j} \rangle &=& V_{a,\alpha_j}=V_{\alpha_j,a}
=\mbox{real} ~,
\eeqa
leading to the Hamiltonian matrix
\beqa\label{matrix}
\left( \begin{array}{cccccc}
E_{a} & V_{a,\alpha_1}     & V_{a,\alpha_2}& V_{a,\alpha_3} &\cdots&V_{a,\alpha_N} \\
V_{a,\alpha_1} & E_{\alpha_1} & 0                        &0                        & \cdots&0          \\
V_{a,\alpha_2} & 0                         &E_{\alpha_2} & 0                       &\cdots&0\\
 . &. &.&.&.\\
V_{a,\alpha_N} & 0                        &0                         &0                          &\cdots&E_{\alpha_N} \\
\end{array} \right) .
\eeqa
Therefore, any eigenstate of the Hamitonian of Eq.(\ref{H}) can be written as
\beqa\label{eigenstate}
| E \rangle = c_{a}(E) | a\rangle
+\sum_{j} c_{\alpha_j}(E)  | \alpha_j\rangle ~.
\eeqa
The above equations and the normalization condition $\langle E \mid E\rangle=1$ lead to the amplitudes
\beqa
c_{\alpha_j}(E)  &=&- c_{a}(E) \frac{ V_{a,\alpha_j} } { (E_{\alpha_j}-E)} \nonumber\\
\left( c_{a}(E)  \right)^2. 
&=&
\left( 1+ \sum_{j} \frac{ \left( V_{ a,\alpha_j } \right)^2}{ ( E_{\alpha_j }-E )^2 }\right)^{-1}~.
\eeqa
Then, the mean value of the energy and the spreading width of the
 state $| a\rangle$ when $E \approx E_a$ are written:
\beqa\label{mean}
\bar E &=& E_a \left(c_a(E)\right)^2 +\sum_j  E_{\alpha_j}\left(c_{\alpha_j}(E)\right)^2. \nonumber\\
\Gamma&=&2\sigma\nonumber\\
&=&2 \left( (E_a - \bar E)^2 \left(c_a(E)\right)^2
+\sum_j  (E_{\alpha_j}-\bar E)^2\left(c_{\alpha_j}(E)\right)^2
\right)^\frac{1}{2}. \nonumber\\
\eeqa

In the next section  we are going to show and discuss the results of our 
calculations, based on the above introduced formalism, both 
for energies and spreading widths of meson-like states. 

\section{Results and Discussion}
\label{results}

In this section we shall analyse the results of our calculations, starting from the energy 
spectrum for each set of quantum numbers of mesonic states.

\begin{figure}[H]
\centering
\includegraphics[width=0.95\textwidth]{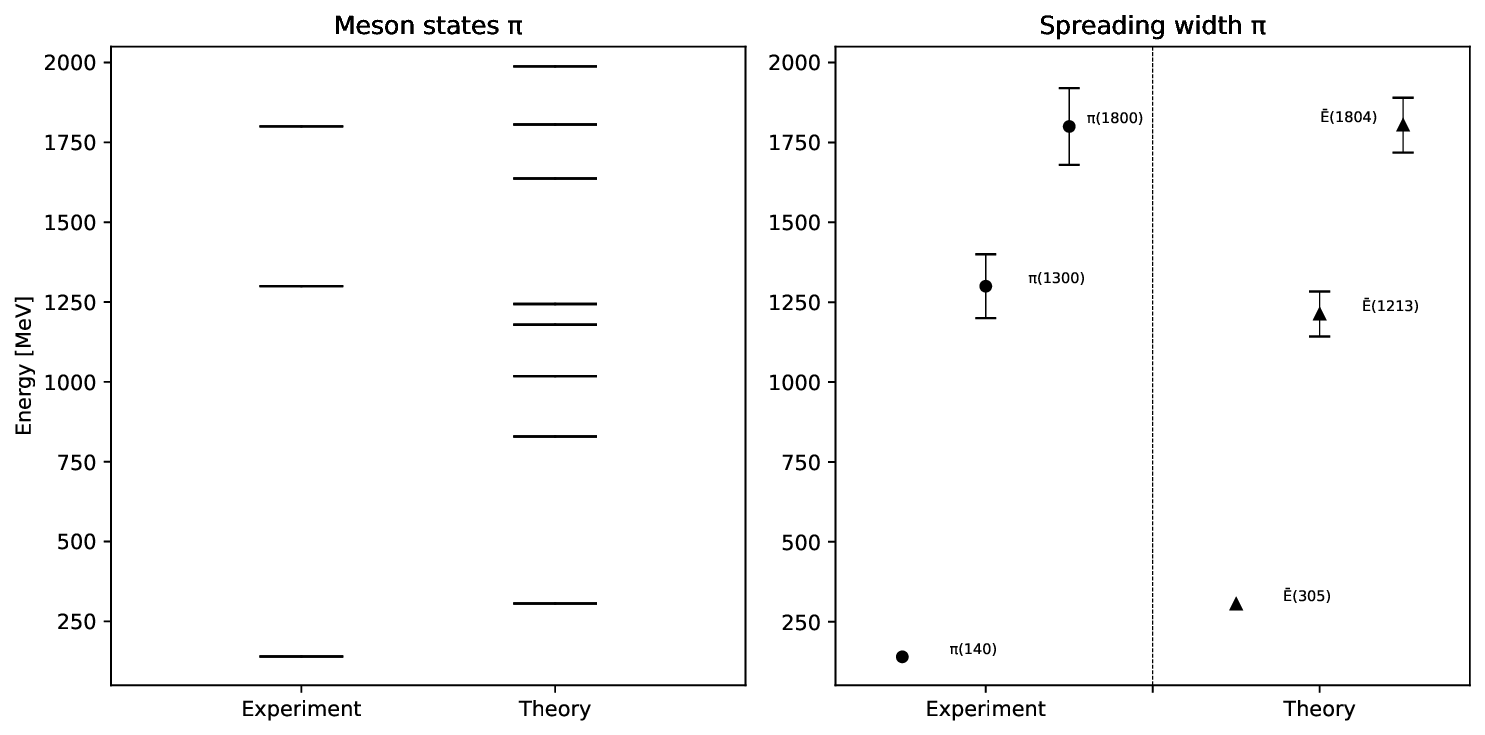}
\caption{Energy (left-side) and spreading width (right-side) of $\pi$ meson states.The theoretical results are compared to the available experimental values\cite{data}.}
\label{fig1}
\end{figure}

In the case of the meson $\pi$ the correspondence between measured states and the theoretically predicted ones is quite acceptable, see Figure \ref{fig1}, for the excited states, both for the energy and for the spreading width of the states, except for the low-lying state where the difference between theory and data is larger than that for the other states. However we shall mention that, in order to determine the validity of our approach, we decided not to adjust the value of the energy of the first $\pi$ state. The energy of the states, shown in the left-hand side of the figure, are then compared to the experimental values after performing with them the calculation of the spreading width (see Eq.(\ref{mean})). 
The strength of the interaction between states was fixed at the value V=70 MeV. This procedure has been applied to all other meson-like subspaces without adjusting the parameters of the model space.

\begin{figure}[H]
\centering
\includegraphics[width=0.95\textwidth]{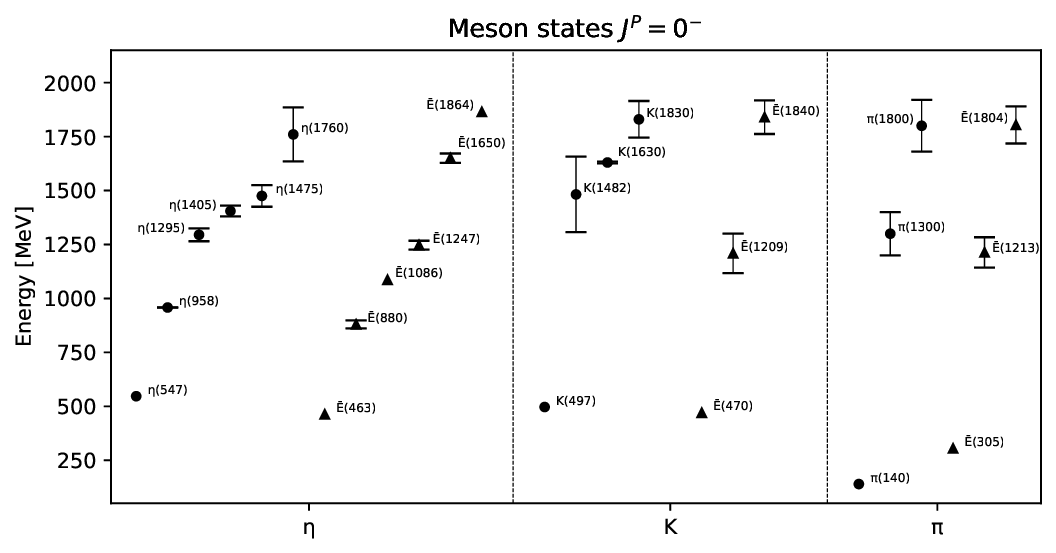}
\caption{Energy spectrum of   $J^{P}=0^-$ meson states. Each of the panels show the experimental values (dots) with the corresponding spreading (vertical lines) and the calculated values (triangles). The centroids of the theoretical values are denoted by $\bar{E}$ and the vertical lines on them are results of the calculated spreading widths}
\label{Fig2}
\end{figure}

The results for the case of the space of mesons with $J^{P}=0^-$
are shown in Figure \ref{Fig2}, where the results corresponding to the meson $\pi$ are included for completeness.

\begin{figure}[H]
\centering
\includegraphics[width=0.95\textwidth]{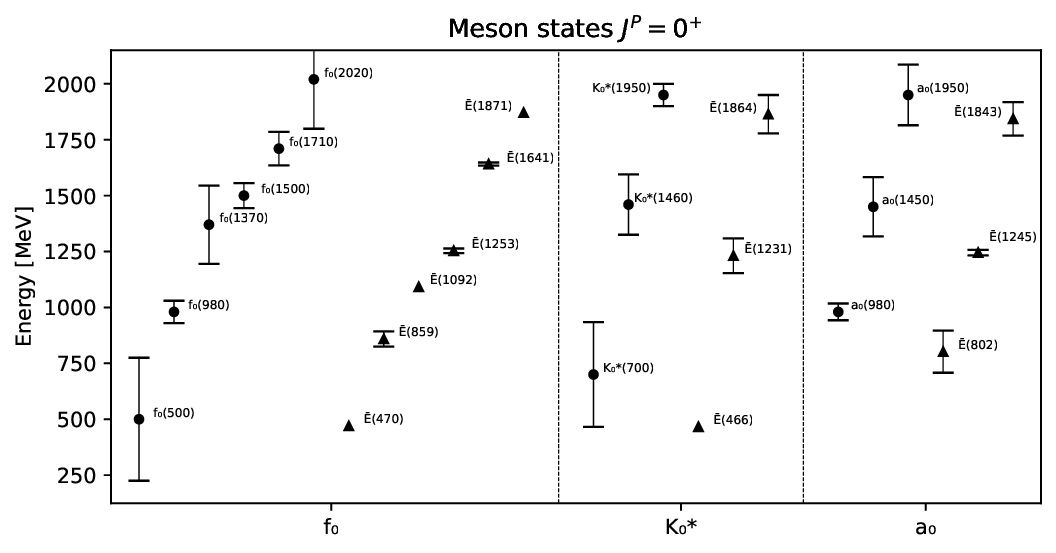}
\caption{Meson states belonging to the $J^{P}=0^+$ space. The experimental and theoretical values are depicted using the notation described in the caption to Fig.(\ref{Fig2}).  }
\label{Fig3}
\end{figure}

\begin{figure}[H]
\centering
\includegraphics[width=0.95\textwidth]{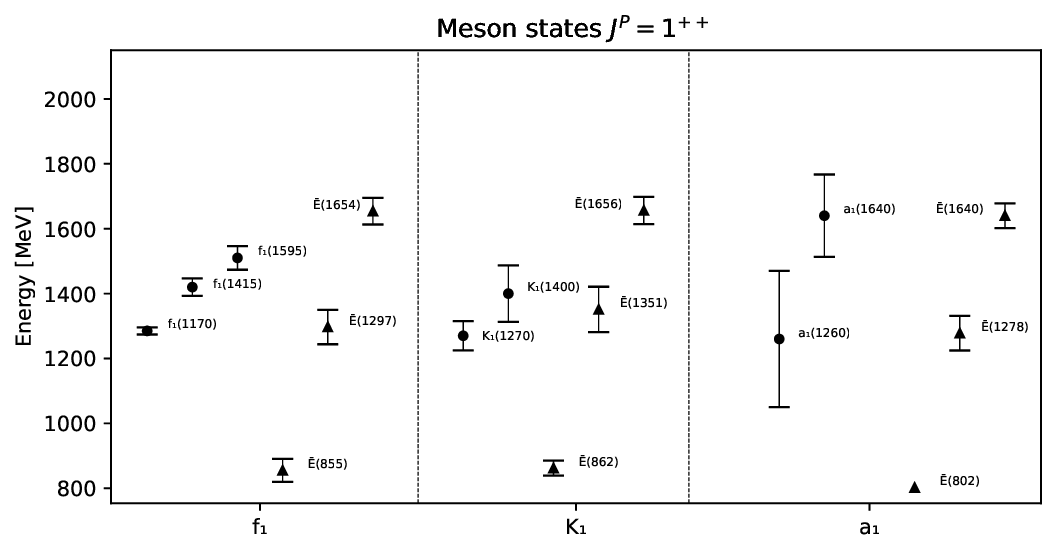}
\caption{ Meson states belonging to the $J^{P}=1^{++}$. See the caption of Fig.(\ref{Fig2}) for the meaning of the symbols .}
\label{Fig4}
\end{figure}

\begin{figure}[H]
\centering
\includegraphics[width=0.95\textwidth]{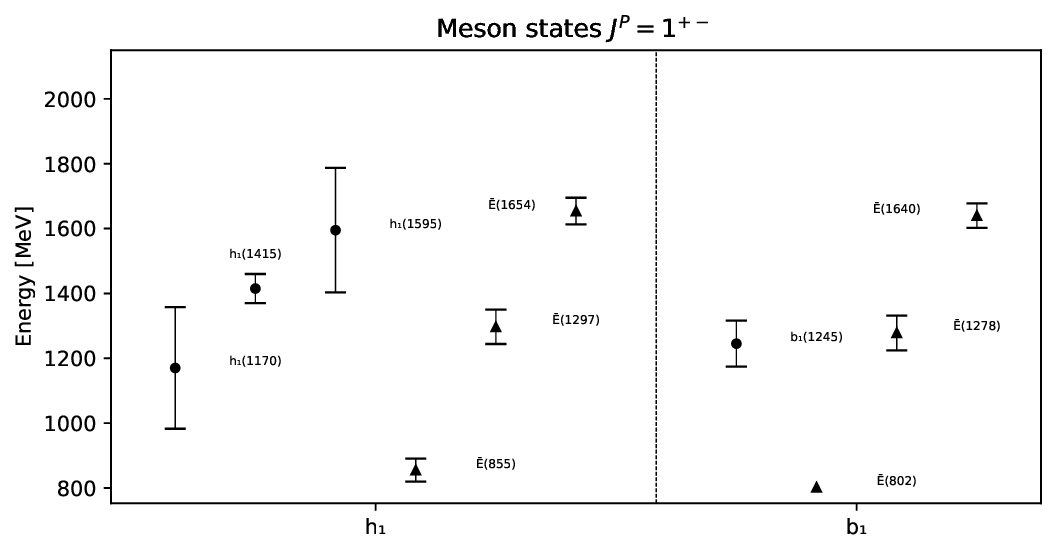}
\caption{Meson states of the  $J^{P}=1^{+-}$ space. The experimental and calculated values are denoted using the symbols introduced in the Fig.(\ref{Fig2}).  }
\label{Fig5}
\end{figure}

\begin{figure}[H]
\centering
\includegraphics[width=0.95\textwidth]{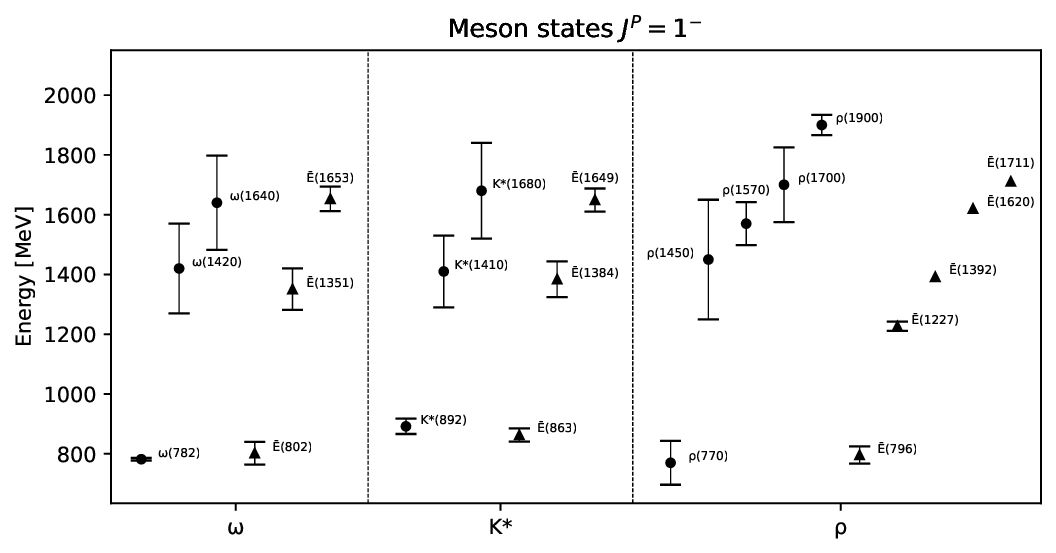}
\caption{Experimental and theoretical results for mesons with $J^{P}=1^{-}$ (See the caption to Fig.(\ref{Fig2})}
\label{Fig6}
\end{figure}

As said before, the calculated low lying state of the pion is shifted to an energy a little bit higher than the experimental one, but since we have taken it as a reference state we did not vary  the parameters of the model in order to better accommodate the result. 
The bunch of states at medium and higher energies can be averaged over and their spreading widths make them compatible with data.  
In the same fashion we are presenting the results for other 
subspaces, namely: mesons with $J^{P}=0^+$ (Fig.(\ref{Fig3})), $J^{P}=1^{++}$ (Fig.(\ref{Fig4})), 
$J^{P}=1^{+-}$(Fig. (\ref{Fig5})), and $J^{P}=1^-$ (Fig.(\ref{Fig6})).

The main discrepancies between data and calculated quantities are found for the cases of the $h_1$ and $K_1 $ mesons 
where the calculation missed the low lying state in both cases and seems to do better at higher energies.
The general trend of the theoretical values is quite acceptable considering that the results have been obtained without adjusting parameters for each mesonic subspace. In all cases we have solved the RPA equations using the same 
space of quasi-quark states to built the configurations restricted by the couplings associated to each mesonic representation.

\section{Conclusions}
\label{conclusions}
In this work we have presented our results for the spectrum of meson states with masses up to 2.00 GeV.

We started from the pre-diagonalization of the kinetic energy sector of the QCD Hamiltonian in a basis of harmonic oscillator states. Effective up and down quark states described in this fashion were then transformed to a basis of correlated pairs by means of the Bogoliubov transformations and by applying the BCS scheme.

We have obtained effective quasi-quarks energies and treated the remaining terms of the Hamiltonian in the framework of the Random Phase Approximation. 
In addition to the mass spectra of mesons we have calculated their spreading widths and the theoretical results agree quite well with the experimental ones, with few exceptions.

From the features exhibited by the results it could be said that the gross structure of the observed meson spectra can be explained by the RPA method when applied to effective quasi-quark degrees of freedom. However, some effects which go beyond the RPA scheme are missing and probably the correct procedure would be to proceed further by coupling the correlated two quasi-quark configurations with an extra quasi-quark, that is by adding the H$_{31}$ and H$_{13}$ terms to the Hamiltonian of Eq.(\ref{124}) in order to further renormalized the effective quasi-quark energy to be used in the subsequent RPA scheme.

The advantage of the present procedure consists, as compared to other theoretical approaches, in the formulation of the non-perturbative QCD regime for meson-like states, as a collective many-body problem where strongly correlated states are described as highly coherent superposition of quasi-quark pairs. 

In this fashion the meson spectra seems to be a nice testing ground for the validity of this sort of non-perturbative 
many-body techniques. As we have shown, the bulk of the interaction between quarks can be written in terms of effective quasi-quark degrees of freedom, and the terms of the transformed Hamiltonian  containing pairs of quasi-quarks  can further on  written in terms of collective variables, that is by means of the RPA transformations.   

The same procedure may be also used to compare calculated and experimental energy states in the baryon sector. In this scheme a three quark state, like a single nucleon, can be described as the result of the coupling between correlated two-quasi-quark states coupled to an extra quasi-quark state. Work is in progress about it. 

\section{Acknowledgements}

P.O.H. acknowledges financial support from PAPIIT-DGAPA
(IN116824). O.C acknowledges the support of the CONICET PIP 11220200102081CO and the PICT17-03-0571
of the ANPCyT of Argentina.

\section{Appendices}
Here, we are listing and commenting upon the expressions which have been utilized in the text, particularly, the components of the Coulomb interaction written in the quasiparticle basis, as well as the expression of the 
different terms of the Hamiltonian in that basis.
\subsection{Appendix A.1}
In this Appendix we are given the expressions of the matrix elements of the Coulomb interaction, as they appear in Eq.(\ref{coulomb}). They are written:
\beqa
 V_{\{N_i l_i j_i\}}^{L}
 &=& 
\sum_{JN^\prime_rN_rl_rN_RL_R}
3\sqrt{8}\sqrt{(2j_1+1)(2j_2+1)(2j_3+1)(2j_4+1)}\sqrt{2L+1}(2J+1)
\nonumber \\
&\times &
(-1)^{L+j_2+j_4-J+1}
\left\{
\begin{array}{ccc}
l_1 & L & l_2 \\
j_2 & \frac{1}{2} & j_1
\end{array}
\right\}
\left\{
\begin{array}{ccc}
l_3 & L & l_4 \\
j_4 & \frac{1}{2} & j_3
\end{array}
\right\}
\left\{
\begin{array}{ccc}
l_2 & J & l_4 \\
l_3 & L & l_1
\end{array}
\right\}
\nonumber \\
&\times&
(  N^\prime_rl_r,N_RL_R,J\mid N_1l_1,N_3l_3,J  )
(  N_rl_r,N_RL_R,J\mid N_2l_2,N_4l_4,J  )
\int d^3r \Psi^*_{N^\prime_r l_r l_r}(\vec{r}) V(\sqrt{2}r)\Psi_{N_rl_rl_r}(\vec{r})
~~~.\nonumber\\
\eeqa
The diagonalization of the kinetic term (\ref{kinetic}) is performed for a given 
maximal number of quanta $N=N_{{\rm cut}}$, for which we introduce a general transformation
to a basis of effective operators,
\beqa
{   q}^\dagger_{\tau (N l)jmcf} & = &
\sum_{\lambda \pi k} \left( \alpha^{j,T}_{\tau (Nl),\lambda \pi k}\right)^* 
%{\hat {   b}}
{   Q}^\dagger_{\lambda \pi kjmcf}  
~\delta_{\pi,(-1)^{\frac{1}{2}-\tau+l}}
~~~.
\label{eqp}
\eeqa

The index $\lambda = \pm\frac{1}{2}$ refers 
to the pseudo-spin components 
after the diagonalization of the kinetic term, and
$k$ runs over all orbital states after  the  diagonalization. 
The value $\lambda = +\frac{1}{2}$ refers to positive energy states
(effective quarks) and the value
$\lambda = -\frac{1}{2}$ to negative energy states (effective antiquarks). 
For example, taking $N_{{\rm cut}}=3$ and
$j=\frac{1}{2}$, only the lowest two $s$ and $p$ orbital states contribute,
so that the index for the
positive energy states runs from $k$ = 1 to 2, for a fixed $\tau$.
The same happens for  the negative energy states.
The parity on the left-hand side (L.H.S) in (\ref{eqp}) is given by
$ (-1)^{\frac{1}{2}-\tau+l}$  while for the right-hand side (R.H.S.)
$\pi$ indicates the parity after the diagonalization {\it i.e.}, our
effective particles $({   Q}^\dagger_{\lambda \pi kjmcf})$ are a
linear combination of states with the same well-defined parity. 

The transformation coefficients depend on the type of quarks, whether it is
an up or down quark
(equal masses are assumed) or a strange quark ($m_{u,d}<m_s$). 
For the transformation coefficients and the kinetic matrix elements,
only the dependence on the flavor isospin is given,
because the flavor-hypercharge is fixed by $T$.

The eigenvalue problem to be solved acquires the following form
\beqa\label{eq:prediag}
\sum_{\tau_i  N_i l_i } 
 \alpha^{j,T}_{\tau_1,(N_1 l_1),   \lambda_1 \pi_1 k_1} 
K^{j,T}_{\tau_1(N_1l_1),\tau_2(N_2l_2)}
 \alpha^{j,T}_{\tau_2,(N_2 l_2), \lambda_2 \pi_2 k_2} 
=\varepsilon_{\lambda_1 \pi_1 k_1 j Y T}
~\delta_{\lambda_1 \lambda_2}\delta_{\pi_1 \pi_2}\delta_{k_1 k_2}~,
\eeqa
where we have taken the transformation coefficients of
Eq.(\ref{eq:prediag}) to be real.

It is worth mentioning that the spin, color and flavor quantum numbers
of the single particles are conserved in the prediagonalization. 
In  this work, we only consider single particles with total spin
$j=\frac{1}{2}$ in order to study low energy states built by pairs coupled to spin $J=0,1$
and leave for future work the analysis of the contributions due to
single particle spin $j=\frac{3}{2}$. The method presented here can directly be extended to $j > \frac{1}{2}$.

Having performed the diagonalization of Eq. (\ref{eq:prediag}), 
the kinetic energy is rewritten in terms of effective
creation (annihilation) operators of 
quarks ${   b}^\dagger({   b}$) and antiquarks ${   d}^\dagger({    d}$):

\beqa
{   K} &   =   &
\sum_{\pi k j Y T} \varepsilon_{\pi k j Y T} 
\sum_{mcT_z}\left(
{   b}^\dagger_{\pi k j Y T, m c T_z} {   b}^{\pi k j Y T,m c T_z}
-
{   d}^{\pi k j Y T, m c T_z}  {   d}^\dagger_{\pi k j Y T, m c T_z}
\right)
~~~,
\eeqa
where the eigenvalues (effective masses) are denoted by
$\varepsilon_{\pi k j Y T}$ and the quark and antiquark operators are related
to the operators before the prediagonalization via 
${   Q}^\dagger_{\frac{1}{2} \pi kjmcf}  \rightarrow  {    b}^\dagger_{\pi k j Y T, m c T_z}$
and
${   Q}^\dagger_{-\frac{1}{2} \pi kjmcf}  \rightarrow  {   d}_{\pi  k j Y T, m c T_z}$.
The diagonalization is achieved by 
choosing the oscillator length $(\sqrt{B_0})^{-1}$ in such a way that for a given cut-off $N_{{\rm cut}}=N_0$,
called the renormalization point,
the first ($k=1$)  single-particle state $\varepsilon_{\pi k j Y T}$
is at a fixed energy. 
This energy is chosen such that it represents a confined particle in a finite volume of the
size of a hadron, {\it i.e.}, $0.5\mbox{fm} \le (\sqrt{B_0})^{-1} \le 1\mbox{fm}$. 
\subsection{Appendix A.2}
The explicit expression of the factors $\mathcal{F}$ and $\mathcal{G}$ of Eq.(\ref{pairs}) in the quasiparticle basis are written:
\beqa
&&\mathcal{F}_{12;\Gamma_0 ~\mu_0}
\mathcal{F}_{34;\bar \Gamma_0 ~\bar \mu_0}\nonumber\\
&&=\frac{1}{2}\sum_{ m_{J_1}m_{J_2}, c_1 c_2, M_{T_1}M_{T_2} } 
\langle J_1 m_1,J_2-m_2\mid LM_L\rangle 
\langle (10)c_1,(01){\bar c}_2\mid (11) C\rangle_1 
\frac{(-1)^{T_1-M_{T_1}}}{\sqrt{2T_1+1}} \delta_{M_{T_1}  M_{T_2}}
(-1)^{J_2-m_{J_2}} (-1)^{\chi_{c_2}}(-1)^{\chi_{f_2}}\nonumber\\
&&\times \sum_{ m_{J_3}m_{J_4}, c_3 c_4, M_{T_3}M_{T_4} } 
\langle J_3 MK3D, J_4-m_4\mid L-M_L\rangle 
\langle (10)c_3,(01){\bar c}_4\mid (11) \bar{C}\rangle_1 
\frac{(-1)^{T_3-M_{T_3}}}{\sqrt{2T_3+1}} \delta_{M_{T_3}  M_{T_4}}
(-1)^{J_4-m_{J_4}} (-1)^{\chi_{c_4}}(-1)^{\chi_{f_4}}\nonumber\\
&&\bigg \{
-(\delta_{++}u_{k_1}u_{k_2} +\delta_{--} v_{k_1} v_{k_2})  ~
(\delta_{++}u_{k_3}u_{k_4} +\delta_{--} v_{k_3} v_{k_4})(  { B}^\dagger_{k_1 \mu_1}   { B}^\dagger_{k_3 \mu_3}
  { B}^{ k_2 \mu_2}     { B}^{ k_4 \mu_4})\nonumber\\
&&
~~ + (\delta_{++}u_{k_1}u_{k_2} +\delta_{--} v_{k_1} v_{k_2})  ~
(\delta_{++}v_{k_3} v_{k_4}+\delta_{--}u_{k_3}u_{k_4})~
(  { B}^\dagger_{k_1 \mu_1}    { D}^{\dagger \ k_4  \mu_4}
  { B}^{ k_2 \mu_2}   { D}_{k_3 \mu_3}) \nonumber\\
&&
~~ +
(\delta_{++}u_{k_1}u_{k_2} +\delta_{--} v_{k_1} v_{k_2})  ~
(\delta_{++}u_{k_3}v_{k_4}-\delta_{--}v_{k_3}u_{k_4})   ~
 (  { B}^\dagger_{k_3 \mu_3}   { D}^{\dagger \ k_4 \mu_4})~
(  { B}^\dagger_{k_1 \mu_1}   { B}^{ k_2 \mu_2})
\nonumber\\
&&
~~ +(\delta_{++}u_{k_1}u_{k_2} +\delta_{--} v_{k_1} v_{k_2})  ~
(\delta_{++}v_{k_3}u_{k_3}-\delta_{--}u_{k_3}v_{k_3})  ~
 (  { B}^\dagger_{k_1 \mu_1}   { B}^{ k_2 \mu_2})~
 (  {  D}_{k_3 \mu_3}   { B}^{ k_4 \mu_4} )\nonumber \\
&&
~~-(\delta_{++}v_{k_1} v_{k_2}+\delta_{--}u_{k_1}u_{k_2})~
(\delta_{++}u_{k_3}u_{k_4} +\delta_{--} v_{k_3} v_{k_4})~
 (  { B}^\dagger_{k_3 \mu_3}  { D}^{\dagger \ k_2  \mu_2}
   { D}_{k_1 \mu_1}   { B}^{ k_4 \mu_4})\nonumber\\
&&
~~ -(\delta_{++}v_{k_1} v_{k_2}+\delta_{--}u_{k_1}u_{k_2})~
(\delta_{++}v_{k_3} v_{k_4}+\delta_{--}u_{k_3}u_{k_4})(  { D}^{\dagger \ k_2  \mu_2}   { D}^{\dagger \ k_4  \mu_4} 
  { D}_{k_1 \mu_1}  { D}_{k_3 \mu_3}) \nonumber\\
&&
~~ -
(\delta_{++}v_{k_1} v_{k_2}+\delta_{--}u_{k_1}u_{k_2})~
(\delta_{++}u_{k_3}v_{k_4}-\delta_{--}v_{k_3}u_{k_4})    ~
 (  { B}^\dagger_{k_3 \mu_3}   { D}^{\dagger \ k_4 \mu_4}) ~
 (  { D}^{\dagger \ k_2  \mu_2}   { D}_{k_1 \mu_1})   
\nonumber\\
&&
~~ -(\delta_{++}v_{k_1} v_{k_2}+\delta_{--}u_{k_1}u_{k_2})~
(\delta_{++}v_{k_3}u_{k_4}-\delta_{--}u_{k_3}v_{k_4})  ~
 (  { D}^{\dagger \ k_2  \mu_2}   { D}_{k_1 \mu_1}) ~
 (  {  D}_{k_3 \mu_3}   { B}^{ k_4 \mu_4} )\nonumber\\
&&
~~+(\delta_{++}u_{k_1}v_{k_2}-\delta_{--}v_{k_1}u_{k_2})   ~
(\delta_{++}u_{k_3}u_{k_4} +\delta_{--} v_{k_3} v_{k_4})~
 (  { B}^\dagger_{k_1 \mu_1}   { D}^{\dagger \ k_2 \mu_2}) ~   
 (  { B}^\dagger_{k_3 \mu_3}   { B}^{ k_4 \mu_4})\nonumber\\
&&
~~-(\delta_{++}u_{k_1}v_{k_2}-\delta_{--}v_{k_1}u_{k_2})   ~
(\delta_{++}v_{k_3} v_{k_4}+\delta_{--}u_{k_3}u_{k_4})~
 (  { B}^\dagger_{k_1 \mu_1}   { D}^{\dagger \ k_2 \mu_2}) ~
 (  { D}^{\dagger \ k_4  \mu_4}   { D}_{k_3 \mu_3}) \nonumber\\
&&
~~ +
(\delta_{++}u_{k_1}v_{k_2}-\delta_{--}v_{k_1}u_{k_2})    ~
(\delta_{++}u_{k_3}v_{k_4}-\delta_{--}v_{k_3}u_{k_4})    ~
 (  { B}^\dagger_{k_1 \mu_1}   { D}^{\dagger \ k_2 \mu_2}) ~
 (  { B}^\dagger_{k_3 \mu_3}   { D}^{\dagger \ k_4 \mu_4})    \nonumber\\
&&
~~+(\delta_{++}u_{k_1}v_{k_2}-\delta_{--}v_{k_1}u_{k_2})   ~
(\delta_{++}v_{k_3}u_{k_4}-\delta_{--}u_{k_3}v_{k_4})  ~
 (  { B}^\dagger_{k_1 \mu_1}   { D}^{\dagger \ k_2 \mu_2}) ~
 (  {  D}_{k_3 \mu_3}   { B}^{ k_4 \mu_4} )\nonumber\\
&&
~~+(\delta_{++}v_{k_1}u_{k_2}-\delta_{--}u_{k_1}v_{k_2})  ~
(\delta_{++}u_{k_3}u_{k_4} +\delta_{--} v_{k_3} v_{k_4})~(  { B}^\dagger_{k_3 \mu_3}  {  D}_{k_1 \mu_1} 
  { B}^{ k_2 \mu_2}   { B}^{ k_4 \mu_4})   \nonumber\\
&&
~~-(\delta_{++}v_{k_1}u_{k_2}-\delta_{--}u_{k_1}v_{k_2})  ~
(\delta_{++}v_{k_3} v_{k_4}+\delta_{--}u_{k_3}u_{k_4})~
 (   { D}^{\dagger \ k_4  \mu_4}   {  D}_{k_1 \mu_1} 
  { B}^{ k_2 \mu_2}   { D}_{k_3 \mu_3}) \nonumber\\
&&
~~ +
(\delta_{++}v_{k_1}u_{k_2}-\delta_{--}u_{k_1}v_{k_2})  ~
(\delta_{++}u_{k_3}v_{k_4}-\delta_{--}v_{k_3}u_{k_4})  ~
 (  { B}^\dagger_{k_3 \mu_3}   { D}^{\dagger \ k_4 \mu_4}) ~
 (  {  D}_{k_1 \mu_1}   { B}^{ k_2 \mu_2} )\nonumber\\
&&
~~+(\delta_{++}v_{k_1}u_{k_2}-\delta_{--}u_{k_1}v_{k_2})  ~
(\delta_{++}v_{k_3}u_{k_4}-\delta_{--}u_{k_3}v_{k_4})   ~
 (  {  D}_{k_1 \mu_1}   { B}^{ k_2 \mu_2} )~
 (  {  D}_{k_3 \mu_3}   { B}^{ k_4 \mu_4} )\nonumber\\
%\bigg \}
&&
+\delta_{23}\bigg\{
\delta_{++++}(u_{k_2}u_{k_3})u_{k_1}u_{k_4}
+\delta_{++--} (u_{k_2} v_{k_3})u_{k_1} v_{k_4}
+\delta_{--++} (v_{k_2} u_{k_3})v_{k_1} u_{k_4}
+\delta_{----}( v_{k_2} v_{k_3}) v_{k_1} v_{k_4}
\bigg\}
  { B}^\dagger_{k_1 \mu_1}   { B}^{ k_4 \mu_4}\nonumber\\
&&
+\delta_{23}
\bigg \{
\delta_{++++}  (u_{k_2} u_{k_3})    u_{k_1}v_{k_4}
-
\delta_{++--}  (u_{k_2} v_{k_3})   u_{k_1}u_{k_4}
+\delta_{--++}  (v_{k_2} u_{k_3})    v_{k_1}v_{k_4}
-
\delta_{----}  (v_{k_2} v_{k_3})   v_{k_1}u_{k_4}
\bigg\}
  {  B}^\dagger_{k_1 \mu_1}  { D}^{\dagger \ k_4 \mu_4}\nonumber\\
&&
+\delta_{14}\bigg\{
\delta_{++++}(v_{k_1} v_{k_4})    v_{k_2}v_{k_3}
+\delta_{++--}(v_{k_1} u_{k_4})   v_{k_2}u_{k_3}
+\delta_{--++}(u_{k_2}v_{k_3})   u_{k_1} v_{k_4}
+\delta_{----}(u_{k_2}u_{k_3})    u_{k_1}u_{k_4}
\bigg\}
  { D}^{\dagger \ k_2  \mu_2}   { D}_{k_3 \mu_3} \nonumber\\
&&
-\delta_{14}
\bigg \{
\delta_{++++}  (v_{k_1} v_{k_4})    v_{k_2}u_{k_3}
-
\delta_{++--}  (v_{k_1} u_{k_4})    v_{k_2}v_{k_3}
+\delta_{--++}  (u_{k_1} v_{k_4})    u_{k_2}u_{k_3}
-
\delta_{----}  (u_{k_1} u_{k_4})    u_{k_2}v_{k_3}
\bigg\}
  { B}^\dagger_{k_3 \mu_3}  { D}^{\dagger \ k_2 \mu_2}\nonumber\\
&&
+\delta_{2 3}\bigg \{ 
\delta_{++++} (u_{k_2}u_{k_3})    v_{k_1}u_{k_4} 
+\delta_{++--} (u_{k_2} v_{k_3})  v_{k_1}v_{k_4}
-\delta_{--++} (v_{k_2}u_{k_3})   u_{k_1}u_{k_4}
-\delta_{----} (v_{k_2} v_{k_3})   u_{k_1}v_{k_4}
\bigg \}
  {  D}_{k_1 \mu_1}   { B}^{ k_4 \mu_4} \nonumber\\
&&
-\delta_{14}\bigg \{ 
\delta_{++++} (v_{k_1} v_{k_4})   u_{k_2}v_{k_3}
+\delta_{++--}(v_{k_1}u_{k_4})   u_{k_2}u_{k_3}
-\delta_{--++}(u_{k_1} v_{k_4})   v_{k_2}v_{k_3}
-\delta_{----}(u_{k_1}u_{k_4})     v_{k_2}u_{k_3}
\bigg \}
  { D}_{k_3 \mu_3}   { B}^{ k_2 \mu_2} 
\nonumber\\
&&
-\delta_{2 3}\bigg \{ 
\delta_{++++}   (u_{k_2} u_{k_3})    v_{k_1}v_{k_4}
-
\delta_{++--}   (u_{k_2} v_{k_3})     v_{k_1}u_{k_4}
-
\delta_{--++}   (v_{k_2} u_{k_3})     u_{k_1}v_{k_4}
+
\delta_{----}   (v_{k_2} v_{k_3})     u_{k_1}u_{k_4}
\bigg\}
  { D}^{\dagger \ k_4 \mu_4}  {  D}_{k_1 \mu_1}\nonumber\\
&&
- \delta_{1 4}\bigg \{ 
\delta_{++++}   (v_{k_1} v_{k_4})    u_{k_2}u_{k_3}
-
\delta_{++--}   (v_{k_1} u_{k_4})     u_{k_2}v_{k_3}
-
\delta_{--++}   (u_{k_1} v_{k_4})     v_{k_2}u_{k_3}
+
\delta_{----}   (u_{k_1} u_{k_4})     v_{k_2}v_{k_3}
\bigg\}
  { B}^\dagger_{k_3 \mu_3}  { B}^{k_2 \mu_2}\nonumber\\
&&
+~
\delta_{1 4} \delta_{2 3}\bigg \{ 
\delta_{++++}   (v_{k_1} v_{k_4})    (u_{k_2}u_{k_3})
-
\delta_{++--}   (v_{k_1} u_{k_4})     (u_{k_2}v_{k_3})
-
\delta_{--++}   (u_{k_1} v_{k_4})     (v_{k_2}u_{k_3})
+
\delta_{----}   (u_{k_1} u_{k_4})     (v_{k_2}v_{k_3})
\bigg\}.\nonumber\\
\eeqa

%FG final
\beqa
&&\mathcal{F}_{12;\Gamma_0 ~\mu_0}
\mathcal{G}_{34;\bar \Gamma_0 ~\bar \mu_0}\nonumber\\
&&=\frac{1}{2}\sum_{ m_{J_1}m_{J_2}, c_1 c_2, M_{T_1}M_{T_2} } 
\langle J_1 m_1,J_2-m_2\mid LM_L\rangle 
\langle (10)c_1,(01){\bar c}_2\mid (11) C\rangle_1 
\frac{(-1)^{T_1-M_{T_1}}}{\sqrt{2T_1+1}} \delta_{M_{T_1}  M_{T_2}}
(-1)^{J_2-m_{J_2}} (-1)^{\chi_{c_2}}(-1)^{\chi_{f_2}}\nonumber\\
&&\times \sum_{ m_{J_3}m_{J_4}, c_3 c_4, M_{T_3}M_{T_4} } 
\langle J_3 m_3, J_4-m_4\mid L-M_L\rangle 
\langle (10)c_3,(01){\bar c}_4\mid (11) \bar{C}\rangle_1 
\frac{(-1)^{T_3-M_{T_3}}}{\sqrt{2T_3+1}} \delta_{M_{T_3}  M_{T_4}}
(-1)^{J_4-m_{J_4}} (-1)^{\chi_{c_4}}(-1)^{\chi_{f_4}}\nonumber\\
&&
\bigg \{
+(\delta_{++}u_{k_1}u_{k_2} +\delta_{--} v_{k_1} v_{k_2}) 
( \delta_{+-}  u_{k_3}v_{k_4} ~+~  \delta_{-+}  v_{k_3} u_{k_4} )
  { B}^\dagger_{k_1 \mu_1}   { B}^\dag_{k_3 \mu_3} 
  { B}^{ k_2 \mu_2}   { B}^{k_4 \mu_4} \nonumber\\
&&
~~~~-(\delta_{++}u_{k_1}u_{k_2} +\delta_{--} v_{k_1} v_{k_2}) 
(  \delta_{-+} u_{k_3}v_{k_4} ~+~  \delta_{+-} v_{k_3} u_{k_4}   )
  { B}^\dagger_{k_1 \mu_1}   { D}^{\dagger \ k_4  \mu_4}
  { D}_{k_3 \mu_3}  { B}^{ k_2 \mu_2} \nonumber\\
&&
~~~~+(\delta_{++}u_{k_1}u_{k_2} +\delta_{--} v_{k_1} v_{k_2}) 
(\delta_{-+} u_{k_3}u_{k_4} ~-~  \delta_{+-} v_{k_3} v_{k_4})    
(  { B}^\dagger_{k_1 \mu_1}   { B}^{ k_2 \mu_2})
(  { D}_{k_3 \mu_3}   { B}^{ k_4 \mu_4})\nonumber\\
&&
~~~~+(\delta_{++}u_{k_1}u_{k_2} +\delta_{--} v_{k_1} v_{k_2}) 
(\delta_{+-} u_{k_3}u_{k_4}  ~-~ \delta_{-+} v_{k_3} v_{k_4} ) 
  { B}^\dag_{k_3\mu_3}   { D}^{\dagger \ k_4\mu_4} 
  { B}^\dagger_{k_1 \mu_1}   { B}^{ k_2 \mu_2}\nonumber\\   
&&
~~~~+(\delta_{++}v_{k_1} v_{k_2}+\delta_{--}u_{k_1}u_{k_2}) 
( \delta_{+-}  u_{k_3}v_{k_4} ~+~  \delta_{-+}  v_{k_3} u_{k_4} )
  { B}^\dag_{k_3 \mu_3}  { D}^{\dagger \ k_2  \mu_2} 
  { D}_{k_1 \mu_1}   { B}^{k_4 \mu_4} \nonumber\\
&&
~~~~-(\delta_{++}v_{k_1} v_{k_2}+\delta_{--}u_{k_1}u_{k_2})  
(  \delta_{-+} u_{k_3}v_{k_4} ~+~  \delta_{+-} v_{k_3} u_{k_4}   )
  { D}^{\dagger \ k_2  \mu_2}   { D}^{\dagger \ k_4  \mu_4}
  { D}_{k_1 \mu_1}  { D}_{k_3 \mu_3}\nonumber\\
&&
~~~~-(\delta_{++}v_{k_1} v_{k_2}+\delta_{--}u_{k_1}u_{k_2}) 
(\delta_{-+} u_{k_3}u_{k_4} ~-~  \delta_{+-} v_{k_3} v_{k_4})    
(  { D}^{\dagger \ k_2  \mu_2}   { D}_{k_1 \mu_1})
(  { D}_{k_3 \mu_3}   { B}^{ k_4 \mu_4})\nonumber\\
&&
~~~~-(\delta_{++}v_{k_1} v_{k_2}+\delta_{--}u_{k_1}u_{k_2}) 
(\delta_{+-} u_{k_3}u_{k_4}  ~-~ \delta_{-+} v_{k_3} v_{k_4} )
(  { B}^\dag_{k_3\mu_3}   { D}^{\dagger \ k_4\mu_4} )
(  { D}^{\dagger \ k_2  \mu_2}   { D}_{k_1 \mu_1})\nonumber\\    
&&
~~~~-(\delta_{++}u_{k_1}v_{k_2}-\delta_{--}v_{k_1}u_{k_2})
( \delta_{+-}  u_{k_3}v_{k_4} ~+~  \delta_{-+}  v_{k_3} u_{k_4} )
(  { B}^\dagger_{k_1 \mu_1}   { D}^{\dagger \ k_2 \mu_2})
 (  { B}^\dag_{k_3 \mu_3}   { B}^{k_4 \mu_4} )\nonumber\\
&&
~~~~-(\delta_{++}u_{k_1}v_{k_2}-\delta_{--}v_{k_1}u_{k_2})
(  \delta_{-+} u_{k_3}v_{k_4} ~+~  \delta_{+-} v_{k_3} u_{k_4}   )
(  { B}^\dagger_{k_1 \mu_1}   { D}^{\dagger \ k_2 \mu_2})
(  { D}^{\dagger \ k_4  \mu_4}   { D}_{k_3 \mu_3})\nonumber\\
&&
~~~~+(\delta_{++}u_{k_1}v_{k_2}-\delta_{--}v_{k_1}u_{k_2})
(\delta_{-+} u_{k_3}u_{k_4} ~-~  \delta_{+-} v_{k_3} v_{k_4})    
(  { B}^\dagger_{k_1 \mu_1}   { D}^{\dagger \ k_2 \mu_2})
(  { D}_{k_3 \mu_3}   { B}^{ k_4 \mu_4})\nonumber\\
&&
~~~~+(\delta_{++}u_{k_1}v_{k_2}-\delta_{--}v_{k_1}u_{k_2})
(\delta_{+-} u_{k_3}u_{k_4}  ~-~ \delta_{-+} v_{k_3} v_{k_4} ) 
(  { B}^\dagger_{k_1 \mu_1}   { D}^{\dagger \ k_2 \mu_2})
(  { B}^\dag_{k_3\mu_3}   { D}^{\dagger \ k_4\mu_4} )\nonumber\\
&&
~~~~-(\delta_{++}v_{k_1}u_{k_2}-\delta_{--}u_{k_1}v_{k_2})  
( \delta_{+-}  u_{k_3}v_{k_4} ~+~  \delta_{-+}  v_{k_3} u_{k_4} )
  { B}^\dag_{k_3 \mu_3}  {  D}_{k_1 \mu_1} 
  { B}^{ k_2 \mu_2}    { B}^{k_4 \mu_4} \nonumber\\
&&
~~~~-(\delta_{++}v_{k_1}u_{k_2}-\delta_{--}u_{k_1}v_{k_2})  
(  \delta_{-+} u_{k_3}v_{k_4} ~+~  \delta_{+-} v_{k_3} u_{k_4}   )
  { D}^{\dagger \ k_4  \mu_4}  {  D}_{k_1 \mu_1} 
  { B}^{ k_2 \mu_2}   { D}_{k_3 \mu_3}\nonumber\\
&&
~~~~+(\delta_{++}v_{k_1}u_{k_2}-\delta_{--}u_{k_1}v_{k_2})  
(\delta_{-+} u_{k_3}u_{k_4} ~-~  \delta_{+-} v_{k_3} v_{k_4})    
(  {  D}_{k_1 \mu_1}   { B}^{ k_2 \mu_2} )
(  { D}_{k_3 \mu_3}   { B}^{ k_4 \mu_4})\nonumber\\
&&
~~~~+(\delta_{++}v_{k_1}u_{k_2}-\delta_{--}u_{k_1}v_{k_2})  
(\delta_{+-} u_{k_3}u_{k_4}  ~-~ \delta_{-+} v_{k_3} v_{k_4} ) 
(  { B}^\dag_{k_3\mu_3}   { D}^{\dagger \ k_4\mu_4} )
(  {  D}_{k_1 \mu_1}   { B}^{ k_2 \mu_2} )\nonumber\\    
&&
-\delta_{23}\bigg\{
\delta_{+++-} (u_{k_2}  u_{k_3})    u_{k_1}v_{k_4}
+\delta_{++-+} (u_{k_2}  v_{k_3})    u_{k_1} u_{k_4}
+\delta_{--+-}( v_{k_2}  u_{k_3})    v_{k_1} v_{k_4}
+\delta_{---+}( v_{k_2}  v_{k_3})     v_{k_1} u_{k_4}\bigg\}
  { B}^\dagger_{k_1\mu_1}  { B}^{k_4\mu_4} \nonumber\\
&&
+\delta_{2 3}
\bigg \{
\delta_{+++-}
(u_{k_2} u_{k_3})    u_{k_1}u_{k_4}
-\delta_{++-+}
(u_{k_2} v_{k_3})    u_{k_1}v_{k_4}
% &&
+\delta_{--+-}
(v_{k_2} u_{k_3})     v_{k_1}u_{k_4}
-\delta_{---+}
(v_{k_2} v_{k_3})    v_{k_1}v_{k_4}\bigg\}
  {  B}^\dagger_{k_1 \mu_1}  { D}^{\dagger \ k_4 \mu_4}\nonumber\\
&&
+\delta_{14}
\bigg\{
\delta_{++-+} (v_{k_1} v_{k_4} )   v_{k_2} u_{k_3}
+\delta_{+++-} (v_{k_1} u_{k_4} )     v_{k_2} v_{k_3}
+\delta_{---+} (u_{k_1}v_{k_4} )      u_{k_2} u_{k_3}
+\delta_{--+-} (u_{k_1} u_{k_4} )    u_{k_2} v_{k_3}
\bigg\}
  { D}^{\dagger \ k_2 \mu_2}  {D}_{k_3\mu_3}\nonumber\\
&&
-\delta_{1 4}
\bigg \{
\delta_{+++-}
(v_{k_1} u_{k_4})    v_{k_2}u_{k_3}
-\delta_{++-+}
(v_{k_1} v_{k_4})    v_{k_2}v_{k_3}
% &&
+\delta_{--+-}
(u_{k_1} u_{k_4})    u_{k_2}u_{k_3}
-\delta_{---+}
(u_{k_1} v_{k_4})    u_{k_2}v_{k_3}\bigg\}
  { B}^\dagger_{k_3 \mu_3}  { D}^{\dagger \ k_2 \mu_2}\nonumber\\
&&
-\delta_{23}\bigg\{
\delta_{+++-} (u_{k_2}  u_{k_3})    v_{k_1}v_{k_4}
+\delta_{++-+} (u_{k_2}  v_{k_3})    v_{k_1} u_{k_4}
-\delta_{--+-} (v_{k_2}  u_{k_3})    u_{k_1}v_{k_4}
-\delta_{---+} (v_{k_2}  v_{k_3})    u_{k_1} u_{k_4} 
\bigg\}
  {  D}_{k_1 \mu_1}   { B}^{k_4 \mu_4} \nonumber\\
&& 
-\delta_{14}\bigg\{
\delta_{++-+} (v_{k_1}v_{k_4})       u_{k_2} u_{k_3}
+\delta_{+++-}(v_{k_1} u_{k_4})    u_{k_2} v_{k_3}
-\delta_{---+}(u_{k_1}v_{k_4})      v_{k_2} u_{k_3}
-\delta_{--+-} (u_{k_1} u_{k_4})    v_{k_2} v_{k_3}
\bigg\}
  { D}_{k_3 \mu_3}  { B}^{ k_2 \mu_2} \nonumber\\
&& 
-\delta_{2 3}\bigg \{ 
\delta_{+++-}   (u_{k_2} u_{k_3})    v_{k_1}u_{k_4}
-
\delta_{++-+}   (u_{k_2} v_{k_3})     v_{k_1}v_{k_4}
-
\delta_{--+-}   (v_{k_2} u_{k_3})     u_{k_1}u_{k_4}
+
\delta_{---+}   (v_{k_2} v_{k_3})     u_{k_1}v_{k_4}
\bigg\}
  { D}^{\dagger \ k_4 \mu_4}  {  D}_{k_1 \mu_1}\nonumber\\
&&
- \delta_{1 4}\bigg \{ 
\delta_{+++-}   (v_{k_1} u_{k_4})    u_{k_2}u_{k_3}
-
\delta_{++-+}   (v_{k_1} v_{k_4})     u_{k_2}v_{k_3}
-
\delta_{--+-}   (u_{k_1} u_{k_4})     v_{k_2}u_{k_3}
+
\delta_{---+}   (u_{k_1} v_{k_4})     v_{k_2}v_{k_3}
\bigg\}
  { B}^\dagger_{k_3 \mu_3}  { B}^{k_2 \mu_2}\nonumber\\
&&
+
\delta_{1 4} \delta_{2 3}\bigg \{ 
\delta_{+++-}   (v_{k_1} u_{k_4})    (u_{k_2}u_{k_3})
-
\delta_{++-+}   (v_{k_1} v_{k_4})     (u_{k_2}v_{k_3})
-
\delta_{--+-}   (u_{k_1} u_{k_4})     (v_{k_2}u_{k_3})
+
\delta_{---+}   (u_{k_1} v_{k_4})     (v_{k_2}v_{k_3})
\bigg\}. \nonumber \\
\eeqa

\beqa
&&\mathcal{G}_{12;\Gamma_0 ~\mu_0}
\mathcal{F}_{34;\bar \Gamma_0 ~\bar \mu_0}\nonumber\\
&&=\frac{1}{2}\sum_{ m_{J_1}m_{J_2}, c_1 c_2, M_{T_1}M_{T_2} } 
\langle J_1 m_1,J_2-m_2\mid LM_L\rangle 
\langle (10)c_1,(01){\bar c}_2\mid (11) C\rangle_1 
\frac{(-1)^{T_1-M_{T_1}}}{\sqrt{2T_1+1}} \delta_{M_{T_1}  M_{T_2}}
(-1)^{J_2-m_{J_2}} (-1)^{\chi_{c_2}}(-1)^{\chi_{f_2}}\nonumber\\
&&\times \sum_{ m_{J_3}m_{J_4}, c_3 c_4, M_{T_3}M_{T_4} } 
\langle J_3 m_3, J_4-m_4\mid L-M_L\rangle 
\langle (10)c_3,(01){\bar c}_4\mid (11) \bar{C}\rangle_1 
\frac{(-1)^{T_3-M_{T_3}}}{\sqrt{2T_3+1}} \delta_{M_{T_3}  M_{T_4}}
(-1)^{J_4-m_{J_4}} (-1)^{\chi_{c_4}}(-1)^{\chi_{f_4}}\nonumber\\
&&
\bigg \{
+( \delta_{+-}  u_{k_1}v_{k_2} ~+~  \delta_{-+}  v_{k_1} u_{k_2} ) 
(\delta_{++}u_{k_3}u_{k_4} +\delta_{--} v_{k_3} v_{k_4})
  { B}^\dagger_{k_1 \mu_1}   { B}^\dag_{k_3 \mu_3}
  { B}^{ k_2 \mu_2}   { B}^{k_4 \mu_4}\nonumber\\
&&
~~~~+( \delta_{+-}  u_{k_1}v_{k_2} ~+~  \delta_{-+}  v_{k_1} u_{k_2} ) 
(\delta_{++}v_{k_3} v_{k_4}+\delta_{--}u_{k_3}u_{k_4}) 
  { B}^\dagger_{k_1 \mu_1}   { D}^{\dagger \ k_4  \mu_4}
  { D}_{k_3 \mu_3}  { B}^{ k_2 \mu_2}\nonumber\\
&&
~~~~-( \delta_{+-}  u_{k_1}v_{k_2} ~+~  \delta_{-+}  v_{k_1} u_{k_2} ) 
(\delta_{++}v_{k_3}u_{k_4}-\delta_{--}u_{k_3}v_{k_4})
(  { B}^\dagger_{k_1 \mu_1}   { B}^{ k_2 \mu_2})
(  { D}_{k_3 \mu_3}   { B}^{ k_4 \mu_4})\nonumber\\
&&
~~~~-( \delta_{+-}  u_{k_1}v_{k_2} ~+~  \delta_{-+}  v_{k_1} u_{k_2} ) 
(\delta_{++}u_{k_3}v_{k_4}-\delta_{--}v_{k_3}u_{k_4})
(  { B}^\dag_{k_3\mu_3}   { D}^{\dagger \ k_4\mu_4} )
(  { B}^\dagger_{k_1 \mu_1}   { B}^{ k_2 \mu_2})\nonumber\\
&&
~~~~-(  \delta_{-+} u_{k_1}v_{k_2} ~+~  \delta_{+-} v_{k_1} u_{k_2}   )
(\delta_{++}u_{k_3}u_{k_4} +\delta_{--} v_{k_3} v_{k_4})
  { B}^\dag_{k_3 \mu_3}  { D}^{\dagger \ k_2  \mu_2} 
  { D}_{k_1 \mu_1}   { B}^{k_4 \mu_4} \nonumber\\
&&
~~~~-(  \delta_{-+} u_{k_1}v_{k_2} ~+~  \delta_{+-} v_{k_1} u_{k_2}   )
(\delta_{++}v_{k_3} v_{k_4}+\delta_{--}u_{k_3}u_{k_4}) 
  { D}^{\dagger \ k_2  \mu_2}   { D}^{\dagger \ k_4
  { D}_{k_1 \mu_1} \mu_4}   { D}_{k_3 \mu_3}\nonumber\\
&&
~~~~-(  \delta_{-+} u_{k_1}v_{k_2} ~+~  \delta_{+-} v_{k_1} u_{k_2}   )
(\delta_{++}v_{k_3}u_{k_4}-\delta_{--}u_{k_3}v_{k_4})
(  { D}^{\dagger \ k_2  \mu_2}   { D}_{k_1 \mu_1})
(  { D}_{k_3 \mu_3}   { B}^{ k_4 \mu_4})\nonumber\\
&&
~~~~-(  \delta_{-+} u_{k_1}v_{k_2} ~+~  \delta_{+-} v_{k_1} u_{k_2}   )
(\delta_{++}u_{k_3}v_{k_4}-\delta_{--}v_{k_3}u_{k_4})
(  { B}^\dag_{k_3\mu_3}   { D}^{\dagger \ k_4\mu_4} )
 (  { D}^{\dagger \ k_2  \mu_2}   { D}_{k_1 \mu_1})\nonumber\\
&&
~~~~+(\delta_{-+} u_{k_1}u_{k_2} ~-~  \delta_{+-} v_{k_1} v_{k_2})    
(\delta_{++}u_{k_3}u_{k_4} +\delta_{--} v_{k_3} v_{k_4})
  { B}^\dag_{k_3 \mu_3}  {  D}_{k_1 \mu_1} 
  { B}^{ k_2 \mu_2}    { B}^{k_4 \mu_4} \nonumber\\
&&
~~~~-(\delta_{-+} u_{k_1}u_{k_2} ~-~  \delta_{+-} v_{k_1} v_{k_2})    
(\delta_{++}v_{k_3} v_{k_4}+\delta_{--}u_{k_3}u_{k_4}) 
  { D}^{\dagger \ k_4  \mu_4}  {  D}_{k_1 \mu_1} 
  { B}^{ k_2 \mu_2}   { D}_{k_3 \mu_3}\nonumber\\
&&
~~~~+(\delta_{-+} u_{k_1}u_{k_2} ~-~  \delta_{+-} v_{k_1} v_{k_2})    
(\delta_{++}v_{k_3}u_{k_4}-\delta_{--}u_{k_3}v_{k_4})
(  {  D}_{k_1 \mu_1}   { B}^{ k_2 \mu_2} )
(  { D}_{k_3 \mu_3}   { B}^{ k_4 \mu_4})\nonumber\\
&&
~~~~+(\delta_{-+} u_{k_1}u_{k_2} ~-~  \delta_{+-} v_{k_1} v_{k_2})    
(\delta_{++}u_{k_3}v_{k_4}-\delta_{--}v_{k_3}u_{k_4})
(  { B}^\dag_{k_3\mu_3}   { D}^{\dagger \ k_4\mu_4} )
 (  {  D}_{k_1 \mu_1}   { B}^{ k_2 \mu_2} )\nonumber\\
&&
~~~~+(\delta_{+-} u_{k_1}u_{k_2}  ~-~ \delta_{-+} v_{k_1} v_{k_2} ) 
(\delta_{++}u_{k_3}u_{k_4} +\delta_{--} v_{k_3} v_{k_4})
(  { B}^\dagger_{k_1 \mu_1}   { D}^{\dagger \ k_2 \mu_2})
 (  { B}^\dag_{k_3 \mu_3}   { B}^{k_4 \mu_4} )\nonumber\\
&&
~~~~-(\delta_{+-} u_{k_1}u_{k_2}  ~-~ \delta_{-+} v_{k_1} v_{k_2} ) 
(\delta_{++}v_{k_3} v_{k_4}+\delta_{--}u_{k_3}u_{k_4}) 
(  { B}^\dagger_{k_1 \mu_1}   { D}^{\dagger \ k_2 \mu_2})
(  { D}^{\dagger \ k_4  \mu_4}   { D}_{k_3 \mu_3})\nonumber\\
&&
~~~~+(\delta_{+-} u_{k_1}u_{k_2}  ~-~ \delta_{-+} v_{k_1} v_{k_2} ) 
(\delta_{++}v_{k_3}u_{k_4}-\delta_{--}u_{k_3}v_{k_4})
(  { B}^\dagger_{k_1 \mu_1}   { D}^{\dagger \ k_2 \mu_2})
(  { D}_{k_3 \mu_3}   { B}^{ k_4 \mu_4})\nonumber\\
&&
~~~~+(\delta_{+-} u_{k_1}u_{k_2}  ~-~ \delta_{-+} v_{k_1} v_{k_2} ) 
(\delta_{++}u_{k_3}v_{k_4}-\delta_{--}v_{k_3}u_{k_4})
(  { B}^\dagger_{k_1 \mu_1}   { D}^{\dagger \ k_2 \mu_2})
(  { B}^\dag_{k_3\mu_3}   { D}^{\dagger \ k_4\mu_4} )\nonumber\\
&&
-\delta_{23}
\bigg\{\delta_{+-++}  (v_{k_2} u_{k_3})      u_{k_1}u_{k_4}
+\delta_{+---}  (v_{k_2} v_{k_3})      u_{k_1}v_{k_4}
+\delta_{-+++} (u_{k_2}u_{k_3})       v_{k_1}u_{k_4}
+\delta_{-+--}  (u_{k_2} v_{k_3})     v_{k_1}v_{k_4}
\bigg\}  { B}^\dagger_{k_1 \mu_1}   { B}^{k_4 \mu_4} \nonumber\\
&&
-\delta_{2 3}
\bigg \{
\delta_{+-++}
(v_{k_2} u_{k_3})    u_{k_1}v_{k_4}
-\delta_{+---}
(v_{k_2} v_{k_3})    u_{k_1}u_{k_4}
% &&
+\delta_{-+++}
(u_{k_2} u_{k_3})     v_{k_1}v_{k_4}
-\delta_{-+--}
(u_{k_2} v_{k_3})    v_{k_1}u_{k_4}\bigg\}
  {  B}^\dagger_{k_1 \mu_1}  { D}^{\dagger \ k_4 \mu_4}\nonumber\\
&&+\delta_{14}
\bigg\{
\delta_{-+++}   (u_{k_1} v_{k_4})     v_{k_2} v_{k_3}
+\delta_{-+--}  (u_{k_1}u_{k_4})     v_{k_2} u_{k_3}
+\delta_{+-++} (v_{k_1} v_{k_4})    u_{k_2}v_{k_3}
+\delta_{+---}   (v_{k_1} u_{k_4})     u_{k_2}u_{k_3}
\bigg\}
  {D}^{\dagger \ k_2\mu_2}  {D}_{k_3\mu_3}\nonumber\\
&&
-\delta_{1 4}
\bigg \{
\delta_{-+++}
(u_{k_1} v_{k_4})    v_{k_2}u_{k_3}
-\delta_{-+--}
(u_{k_1} u_{k_4})    v_{k_2}v_{k_3}
% &&
+\delta_{+-++}
(v_{k_1} v_{k_4})    u_{k_2}u_{k_3}
-\delta_{+---}
(v_{k_1} u_{k_4})    u_{k_2}v_{k_3}\bigg\}
  { B}^\dagger_{k_3 \mu_3}  { D}^{\dagger \ k_2 \mu_2}\nonumber\\
&&
+\delta_{23}
\bigg\{\delta_{-+++} (u_{k_2} u_{k_3})    u_{k_1}u_{k_4}
+\delta_{-+--} ( u_{k_2} v_{k_3})   u_{k_1} v_{k_4}
-\delta_{+-++} (v_{k_2}u_{k_3})    v_{k_1} u_{k_4} 
-\delta_{+---} (v_{k_2} v_{k_3})    v_{k_1} v_{k_4} 
\bigg\}  {  D}_{k_1 \mu_1}   { B}^{k_4 \mu_4} \nonumber\\
&& 
-\delta_{14}\bigg\{\delta_{-+++} (u_{k_1} v_{k_4})u_{k_2} v_{k_3}
+\delta_{-+--} (u_{k_1}u_{k_4})     u_{k_2} u_{k_3}
-\delta_{+-++} (v_{k_1} v_{k_4})    v_{k_2}v_{k_3}
-\delta_{+---} (v_{k_1} u_{k_4})     v_{k_2}u_{k_3}
\bigg\}  { D}_{k_3 \mu_3}  { B}^{ k_2 \mu_2}\nonumber\\
&&
-\delta_{2 3}\bigg \{ 
\delta_{-+++}   (u_{k_2} u_{k_3})  ~  u_{k_1}v_{k_4}
-
\delta_{-+--}   (u_{k_2} v_{k_3})  ~   u_{k_1}u_{k_4}
-
\delta_{+-++}   (v_{k_2} u_{k_3})  ~   v_{k_1}v_{k_4}
+
\delta_{+---}   (v_{k_2} v_{k_3})  ~   v_{k_1}u_{k_4}
\bigg\}  { D}^{\dagger \ k_4 \mu_4}  {  D}_{k_1 \mu_1}\nonumber\\
&&
- \delta_{1 4}\bigg \{ 
\delta_{-+++}   (u_{k_1}v_{k_4})  ~  u_{k_2} u_{k_3}
-
\delta_{-+--}   (u_{k_1}u_{k_4})  ~   u_{k_2} v_{k_3}
-
\delta_{+-++}   (v_{k_1}v_{k_4})  ~   v_{k_2} u_{k_3}
+
\delta_{+---}   (v_{k_1}u_{k_4})  ~   v_{k_2} v_{k_3}
\bigg\}
  { B}^\dagger_{k_3 \mu_3}  { B}^{k_2 \mu_2}\nonumber\\
&&
+
\delta_{1 4} \delta_{2 3}\bigg \{ 
\delta_{-+++}   (u_{k_1}v_{k_4})  ~  (u_{k_2} u_{k_3})
-
\delta_{-+--}   (u_{k_1}u_{k_4})  ~   (u_{k_2} v_{k_3})
-
\delta_{+-++}   (v_{k_1}v_{k_4})  ~  ( v_{k_2} u_{k_3})
+
\delta_{+---}   (v_{k_1}u_{k_4})  ~   (v_{k_2} v_{k_3})
\bigg\}. \nonumber\\
\eeqa

and
\beqa
&&\mathcal{G}_{12;\Gamma_0 ~\mu_0}
\mathcal{G}_{34;\bar \Gamma_0 ~\bar \mu_0}\nonumber\\
&&=\frac{1}{2}\sum_{ m_{J_1}m_{J_2}, c_1 c_2, M_{T_1}M_{T_2} } 
\langle J_1 m_1,J_2-m_2\mid LM_L\rangle 
\langle (10)c_1,(01){\bar c}_2\mid (11) C\rangle_1 
\frac{(-1)^{T_1-M_{T_1}}}{\sqrt{2T_1+1}} \delta_{M_{T_1}  M_{T_2}}
(-1)^{J_2-m_{J_2}} (-1)^{\chi_{c_2}}(-1)^{\chi_{f_2}}\nonumber\\
&&\times \sum_{ m_{J_3}m_{J_4}, c_3 c_4, M_{T_3}M_{T_4} } 
\langle J_3 m_3, J_4-m_4\mid L-M_L\rangle 
\langle (10)c_3,(01){\bar c}_4\mid (11) \bar{C}\rangle_1 
\frac{(-1)^{T_3-M_{T_3}}}{\sqrt{2T_3+1}} \delta_{M_{T_3}  M_{T_4}}
(-1)^{J_4-m_{J_4}} (-1)^{\chi_{c_4}}(-1)^{\chi_{f_4}}\nonumber\\
&&
\bigg \{
-( \delta_{+-}  u_{k_1}v_{k_2} ~+~  \delta_{-+}  v_{k_1} u_{k_2} ) 
( \delta_{+-}  u_{k_3}v_{k_4} ~+~  \delta_{-+}  v_{k_3} u_{k_4} ) 
  { B}^\dagger_{k_1 \mu_1}   { B}^\dag_{k_3 \mu_3}
  { B}^{ k_2 \mu_2}   { B}^{k_4 \mu_4} \nonumber\\
&&
~~~~+( \delta_{+-}  u_{k_1}v_{k_2} ~+~  \delta_{-+}  v_{k_1} u_{k_2} ) 
 (  \delta_{-+} u_{k_3}v_{k_4} ~+~  \delta_{+-} v_{k_3} u_{k_4}   )
  { B}^\dagger_{k_1 \mu_1}  { D}^{\dagger \ k_4  \mu_4}
  { D}_{k_3 \mu_3}   { B}^{ k_2 \mu_2}\nonumber\\
&&
~~~~-( \delta_{+-}  u_{k_1}v_{k_2} ~+~  \delta_{-+}  v_{k_1} u_{k_2} ) 
(\delta_{-+} u_{k_3}u_{k_4} ~-~  \delta_{+-} v_{k_3} v_{k_4})    
(  { B}^\dagger_{k_1 \mu_1}   { B}^{ k_2 \mu_2})
(  { D}_{k_3 \mu_3}   { B}^{ k_4 \mu_4})\nonumber\\
&&
~~~~-( \delta_{+-}  u_{k_1}v_{k_2} ~+~  \delta_{-+}  v_{k_1} u_{k_2} ) 
(\delta_{+-} u_{k_3}u_{k_4}  ~-~ \delta_{-+} v_{k_3} v_{k_4} ) 
  { B}^\dag_{k_3\mu_3}   { D}^{\dagger \ k_4\mu_4}
  { B}^\dagger_{k_1 \mu_1}   { B}^{ k_2 \mu_2}\nonumber\\
&&
~~~~+(  \delta_{-+} u_{k_1}v_{k_2} ~+~  \delta_{+-} v_{k_1} u_{k_2}   )
( \delta_{+-}  u_{k_3}v_{k_4} ~+~  \delta_{-+}  v_{k_3} u_{k_4} ) 
  { B}^\dag_{k_3 \mu_3}  { D}^{\dagger \ k_2  \mu_2} 
  { D}_{k_1 \mu_1}  { B}^{k_4 \mu_4} \nonumber\\
&&
~~~~-(  \delta_{-+} u_{k_1}v_{k_2} ~+~  \delta_{+-} v_{k_1} u_{k_2}   )
 (  \delta_{-+} u_{k_3}v_{k_4} ~+~  \delta_{+-} v_{k_3} u_{k_4}   )
  { D}^{\dagger \ k_2  \mu_2}   { D}^{\dagger \ k_4  \mu_4}
  { D}_{k_1 \mu_1}  { D}_{k_3 \mu_3}\nonumber\\
&&
~~~~-(  \delta_{-+} u_{k_1}v_{k_2} ~+~  \delta_{+-} v_{k_1} u_{k_2}   )
(\delta_{-+} u_{k_3}u_{k_4} ~-~  \delta_{+-} v_{k_3} v_{k_4})    
(  { D}^{\dagger \ k_2  \mu_2}   { D}_{k_1 \mu_1})
(  { D}_{k_3 \mu_3}   { B}^{ k_4 \mu_4})\nonumber\\
&&
~~~~-(  \delta_{-+} u_{k_1}v_{k_2} ~+~  \delta_{+-} v_{k_1} u_{k_2}   )
(\delta_{+-} u_{k_3}u_{k_4}  ~-~ \delta_{-+} v_{k_3} v_{k_4} ) 
   { B}^\dag_{k_3\mu_3}   { D}^{\dagger \ k_4\mu_4}
  { D}^{\dagger \ k_2  \mu_2}   { D}_{k_1 \mu_1} \nonumber\\
&&
~~~~-(\delta_{-+} u_{k_1}u_{k_2} ~-~  \delta_{+-} v_{k_1} v_{k_2})    
( \delta_{+-}  u_{k_3}v_{k_4} ~+~  \delta_{-+}  v_{k_3} u_{k_4} ) 
  { B}^\dag_{k_3 \mu_3}  {  D}_{k_1 \mu_1} 
  { B}^{ k_2 \mu_2}   { B}^{k_4 \mu_4} \nonumber\\
&&
~~~~-(\delta_{-+} u_{k_1}u_{k_2} ~-~  \delta_{+-} v_{k_1} v_{k_2})    
 (  \delta_{-+} u_{k_3}v_{k_4} ~+~  \delta_{+-} v_{k_3} u_{k_4}   )
  { D}^{\dagger \ k_4  \mu_4}  {  D}_{k_1 \mu_1} 
  { B}^{ k_2 \mu_2}   { D}_{k_3 \mu_3}\nonumber\\
&&
~~~~+(\delta_{-+} u_{k_1}u_{k_2} ~-~  \delta_{+-} v_{k_1} v_{k_2})    
(\delta_{-+} u_{k_3}u_{k_4} ~-~  \delta_{+-} v_{k_3} v_{k_4})    
(  {  D}_{k_1 \mu_1}   { B}^{ k_2 \mu_2} )
(  { D}_{k_3 \mu_3}   { B}^{ k_4 \mu_4})\nonumber\\
&&
~~~~+(\delta_{-+} u_{k_1}u_{k_2} ~-~  \delta_{+-} v_{k_1} v_{k_2})    
(\delta_{+-} u_{k_3}u_{k_4}  ~-~ \delta_{-+} v_{k_3} v_{k_4} ) 
  { B}^\dag_{k_3\mu_3}   { D}^{\dagger \ k_4\mu_4}
  {  D}_{k_1 \mu_1}   { B}^{ k_2 \mu_2}\nonumber\\
&&
~~~~-(\delta_{+-} u_{k_1}u_{k_2}  ~-~ \delta_{-+} v_{k_1} v_{k_2} ) 
( \delta_{+-}  u_{k_3}v_{k_4} ~+~  \delta_{-+}  v_{k_3} u_{k_4} ) 
(  { B}^\dagger_{k_1 \mu_1}   { D}^{\dagger \ k_2 \mu_2})
 (  { B}^\dag_{k_3 \mu_3}   { B}^{k_4 \mu_4} )\nonumber\\
&&
~~~~-(\delta_{+-} u_{k_1}u_{k_2}  ~-~ \delta_{-+} v_{k_1} v_{k_2} ) 
 (  \delta_{-+} u_{k_3}v_{k_4} ~+~  \delta_{+-} v_{k_3} u_{k_4}   )
(  { B}^\dagger_{k_1 \mu_1}   { D}^{\dagger \ k_2 \mu_2})
(  { D}^{\dagger \ k_4  \mu_4}   { D}_{k_3 \mu_3})\nonumber\\
&&
~~~~+(\delta_{+-} u_{k_1}u_{k_2}  ~-~ \delta_{-+} v_{k_1} v_{k_2} ) 
(\delta_{-+} u_{k_3}u_{k_4} ~-~  \delta_{+-} v_{k_3} v_{k_4})    
(  { B}^\dagger_{k_1 \mu_1}   { D}^{\dagger \ k_2 \mu_2})
(  { D}_{k_3 \mu_3}   { B}^{ k_4 \mu_4})\nonumber\\
&&
~~~~+(\delta_{+-} u_{k_1}u_{k_2}  ~-~ \delta_{-+} v_{k_1} v_{k_2} ) 
(\delta_{+-} u_{k_3}u_{k_4}  ~-~ \delta_{-+} v_{k_3} v_{k_4} ) 
(  { B}^\dagger_{k_1 \mu_1}   { D}^{\dagger \ k_2 \mu_2})
(  { B}^\dag_{k_3\mu_3}   { D}^{\dagger \ k_4\mu_4} )\nonumber\\
&&
+\delta_{23}
\bigg\{
\delta_{+-+-}  (v_{k_2}  u_{k_3})     u_{k_1}v_{k_4}
+\delta_{+--+}  (v_{k_2}  v_{k_3})    u_{k_1} u_{k_4} 
+ \delta_{-++-}  (u_{k_2}  u_{k_3})    v_{k_1} v_{k_4}
+ \delta_{-+-+}  (u_{k_2}  v_{k_3})    v_{k_1} u_{k_4} 
\bigg\}
  { B}^\dagger_{k_1 \mu_1}  { B}^{k_4 \mu_4} \nonumber\\
&&
-\delta_{2 3}
\bigg \{
\delta_{+-+-}
(v_{k_2} u_{k_3})    u_{k_1}u_{k_4}
-\delta_{+--+}
(v_{k_2} v_{k_3})    u_{k_1}v_{k_4}
% &&
+\delta_{-++-}
(u_{k_2} u_{k_3})     v_{k_1}u_{k_4}
-\delta_{-+-+}
(u_{k_2} v_{k_3})    v_{k_1}v_{k_4}\bigg\}
  {  B}^\dagger_{k_1 \mu_1}  { D}^{\dagger \ k_4 \mu_4}\nonumber\\
&&
+\delta_{14}
\bigg\{\delta_{-+-+}  (u_{k_1}v_{k_4} )    v_{k_2} u_{k_3}
+\delta_{-++-} (u_{k_1} u_{k_4} )    v_{k_2} v_{k_3}
+\delta_{+--+} (v_{k_1} v_{k_4})    u_{k_2} u_{k_3}
+\delta_{+-+-} (v_{k_1} u_{k_4})    u_{k_2} v_{k_3}
\bigg\}
  {D}^{\dagger \ k_2\mu_2}  { D}_{k_3 \mu_3}\nonumber\\
&&
-\delta_{1 4}
\bigg \{
\delta_{-++-}
(u_{k_1} u_{k_4})    v_{k_2}u_{k_3}
-\delta_{-+-+}
(u_{k_1} v_{k_4})    v_{k_2}v_{k_3}
% &&
+\delta_{+-+-}
(v_{k_1} u_{k_4})    u_{k_2}u_{k_3}
-\delta_{+--+}
(v_{k_1} v_{k_4})    u_{k_2}v_{k_3}\bigg\}
  { B}^\dagger_{k_3 \mu_3}  { D}^{\dagger \ k_2 \mu_2}\nonumber\\
&&-\delta_{23}
\bigg\{\delta_{-++-} (u_{k_2}  u_{k_3})    u_{k_1}v_{k_4}
+\delta_{-+-+} (u_{k_2}  v_{k_3})    u_{k_1} u_{k_4}
-\delta_{+-+-} (v_{k_2}  u_{k_3})    v_{k_1} v_{k_4}
-\delta_{+--+} (v_{k_2}  v_{k_3})    v_{k_1} u_{k_4}
\bigg\}
  {  D}_{k_1 \mu_1}   { B}^{k_4 \mu_4} \nonumber\\
&& -\delta_{14}
\bigg\{
\delta_{-+-+} (u_{k_1}v_{k_4})     u_{k_2} u_{k_3}
+\delta_{-++-} (u_{k_1} u_{k_4})    u_{k_2}  v_{k_3}
-\delta_{+--+} (v_{k_1} v_{k_4})     v_{k_2} u_{k_3}
-\delta_{+-+-} (v_{k_1} u_{k_4})    v_{k_2} v_{k_3}
\bigg\}
  { D}_{k_3 \mu_3}  { B}^{ k_2 \mu_2}\nonumber\\
&& 
-\delta_{2 3}\bigg \{ 
\delta_{-++-}   (u_{k_2} u_{k_3})  ~  u_{k_1}u_{k_4}
-
\delta_{-+-+}   (u_{k_2} v_{k_3})  ~   u_{k_1}v_{k_4}
-
\delta_{+-+-}   (v_{k_2} u_{k_3})  ~   v_{k_1}u_{k_4}
+
\delta_{+--+}   (v_{k_2} v_{k_3})  ~   v_{k_1}v_{k_4}
\bigg\}
  { D}^{\dagger \ k_4 \mu_4}  {  D}_{k_1 \mu_1}\nonumber\\
&&
- \delta_{1 4}\bigg \{ 
\delta_{-++-}   (u_{k_1}u_{k_4})  ~  u_{k_2} u_{k_3}
-
\delta_{-+-+}   (u_{k_1}v_{k_4})  ~   u_{k_2} v_{k_3}
-
\delta_{+-+-}   (v_{k_1}u_{k_4})  ~   v_{k_2} u_{k_3}
+
\delta_{+--+}   (v_{k_1}v_{k_4})  ~   v_{k_2} v_{k_3}
\bigg\}
  { B}^\dagger_{k_3 \mu_3}  { B}^{k_2 \mu_2}\nonumber\\
&&
+
\delta_{1 4} \delta_{2 3}\bigg \{ 
\delta_{-++-}   (u_{k_1}u_{k_4})  ~  (u_{k_2} u_{k_3})
-
\delta_{-+-+}   (u_{k_1}v_{k_4})  ~  ( u_{k_2} v_{k_3})
-
\delta_{+-+-}   (v_{k_1}u_{k_4})  ~   (v_{k_2} u_{k_3})
+
\delta_{+--+}   (v_{k_1}v_{k_4})  ~   (v_{k_2} v_{k_3})
\bigg\}. \nonumber\\
\eeqa

\end{document}